\RequirePackage{fix-cm}
\documentclass{svjour3}                     
\smartqed  
\usepackage{graphicx}
\usepackage{natbib}
\usepackage{tabularx}
\usepackage{rotating}
\usepackage{textcomp}
\usepackage{amsmath}
\usepackage{amssymb}
\usepackage{pdfpages}
\begin{document}

\title{The Interplay Between Collisionless Magnetic Reconnection and Turbulence}

\titlerunning{The Interplay Between Collisionless Magnetic Reconnection and Turbulence}

\author{J.~E.~Stawarz \and P.~A.~Mu{\~{n}}oz \and N.~Bessho \and R.~Bandyopadhyay \and T.~K.~M.~Nakamura \and S.~Eriksson \and D.~Graham \and J.~B\"uchner \and A.~Chasapis \and J.~Drake \and M.~A.~Shay \and R.~E.~Ergun \and H.~Hasegawa \and Yu.~V.~Khotyaintsev \and M.~Swisdak \and  F.~Wilder}

\authorrunning{J. E. Stawarz et al.}

\institute{J. E. Stawarz \at
              Department of Mathematics, Physics, and Electrical Engineering, Northumbria University,\\
              Ellison Building, Newcastle upon Tyne NE1 8ST, United Kingdom.\\
              \email{julia.stawarz@northumbria.ac.uk}          
              \and              
              P.~A.~Mu{\~{n}}oz \at             
              Center for Astronomy and Astrophysics, Technical University of Berlin, 10623 Berlin, Germany. \\
              Max Planck Institute for Solar System Research, 37077 Göttingen, Germany. 
              \and           
              N.~Bessho \at
              Department of Astronomy, University of Maryland, College Park, MD 20742, USA.\\
              NASA Goddard Space Flight Center, Greenbelt, MD 20771, USA.           
              \and
              R.~Bandyopadhyay \at
              Department of Astrophysical Sciences, Princeton University, Princeton, NJ 08544, USA.                            
              \and
              T.~K.~M.~Nakamura \at
              Space Research Institute, Austrian Academy of Sciences, 8042 Graz, Austria. \\
              Krimgen LLC, Hiroshima, 7320828, Japan.             
              \and
              S.~Eriksson \at
              Laboratory for Atmospheric and Space Physics, University of Colorado, Boulder, CO, USA.              
              \and
              D.~Graham \at
              Swedish Institute of Space Physics, Uppsala, Sweden.             
              \and 
              J.~B\"uchner \at
              Max Planck Institute for Solar System Research, 37077 Göttingen, Germany. \\
              Center for Astronomy and Astrophysics, Technical University Berlin, 10623 Berlin, Germany.            
              \and
              A.~Chasapis \at
              Laboratory for Atmospheric and Space Physics, Boulder, CO, USA.             
              \and 
              J.~Drake \at
              Institute for Research in Electronics and Applied Physics, University of Maryland, College Park, MD 20740, USA.\\
              Department of Physics, the Institute for Physical Science and Technology and the Joint Space Institute, University of Maryland, College Park, MD 20740, USA.             
              \and
              M.~A.~Shay \at
              Department of Physics and Astronomy, University of Delaware, Newark, Delaware 19716, USA.              
              \and
              R.~E.~Ergun \at
              Laboratory for Atmospheric and Space Physics, University of Colorado at Boulder, Boulder, CO, USA.\\
              Department of Astrophysical and Planetary Sciences, University of Colorado at Boulder, Boulder, CO, USA.              
              \and 
              H.~Hasegawa \at
              Institute of Space and Astronautical Science, JAXA, Sagamihara, Japan.              
              \and
              Yu.~V.~Khotyaintsev \at
              IRF Swedish Institute of Space Physics, Uppsala, Sweden.            
              \and 
              M.~Swisdak \at
              Institute for Research in Electronics and Applied Physics, University of Maryland, College Park, MD, USA.         
              \and 
             F.~Wilder \at              
             University of Texas at Arlington, Arlington, TX, USA.                                                    
}
              
\date{Received: date / Accepted: date}

\maketitle

\begin{abstract} 
Alongside magnetic reconnection, turbulence is another fundamental nonlinear plasma phenomenon that plays a key role in energy transport and conversion in space and astrophysical plasmas. 
From a numerical, theoretical, and observational point of view there is a long history of exploring the interplay between these two phenomena in space plasma environments; however, recent high-resolution, multi-spacecraft observations have ushered in a new era of understanding this complex topic. 
The interplay between reconnection and turbulence is both complex and multifaceted, and can be viewed through a number of different interrelated lenses - including turbulence acting to generate current sheets that undergo magnetic reconnection ({\it turbulence-driven reconnection}), magnetic reconnection driving turbulent dynamics in an environment ({\it reconnection-driven turbulence}) or acting as an intermediate step in the excitation of turbulence, and the random diffusive/dispersive nature of magnetic field lines embedded in turbulent fluctuations enabling so-called {\it stochastic reconnection}. 
In this paper, we review the current state of knowledge on these different facets of the interplay between turbulence and reconnection in the context of collisionless plasmas, such as those found in many near-Earth astrophysical environments, from a theoretical, numerical, and observational perspective.
Particular focus is given to several key regions in Earth's magnetosphere - Earth's magnetosheath, magnetotail, and Kelvin-Helmholtz vortices on the magnetopause flanks - where NASA's {\it Magnetospheric Multiscale} mission has been providing new insights on the topic. 

\keywords{Magnetic Reconnection, Turbulence, Collisionless Plasmas}
\end{abstract}

\section{Introduction}
Many natural plasmas where magnetic reconnection occurs have wide scale separations between the length scales where energy is injected into the system through dynamical processes and the smaller scales where energy is most effectively dissipated. 
Such systems are conducive to the excitation of complex and highly-nonlinear fluctuations, known as turbulence, that transfer energy across scales facilitating the dissipation of large-scale free energy. 

While the basic physics underpinning individual reconnection sites can be considered using idealized models, magnetic reconnection is fundamentally a nonlinear and multi-scale process that both influences and is influenced by the turbulent dynamics in the surrounding plasma. 
Therefore, a complete picture of the onset, development, and interaction of magnetic reconnection with the surrounding plasma requires coupling it into the turbulent dynamics. 
The study of this interaction between turbulence and reconnection has a long history; however, recent high-resolution space plasma observations, notably from NASA's {\it Magnetospheric Multiscale} ({\it MMS}) mission, have allowed us to observationally examine this interaction in greater detail than ever before. 

In this review, we discuss our current understanding of the interaction between magnetic reconnection and turbulence -- particularly within the nearly collisionless plasma regime applicable to many space and astrophysical systems -- from an observational, numerical, and theoretical perspective. 
Sec.~\ref{sec:TurbulentRecon} provides an overview of the varied ways through which turbulence and reconnection can influence each other. 
Sec.~\ref{sec:TurbulentTheory} provides an introduction to elements of turbulence theory geared toward those that may be less familiar with the statistical theory of turbulence. 
Section~\ref{sec:TurbulenceDrivenReconnection} discusses {\it turbulence-driven reconnection} in which reconnection occurs at thin current sheets created by the turbulent dynamics, with Earth's magnetosheath highlighted as a key example where recent progress has been made. 
Sec.~\ref{sec:ReconnectionDrivenTurbulence} discusses {\it reconnection-driven turbulence} in which reconnection at a pre-existing current sheet excites turbulence, with Earth's magnetotail highlighted as a key region of recent progress.
Sec.~\ref{sec:KHI} discusses magnetic reconnection as an element in the process of large-scale structures transitioning to a turbulent state, with a focus on the Kelvin-Helmholtz Instability (KHI) on Earth's magnetopause. 
Sec.~\ref{sec:StocasticReconnection} discusses how the stochastic nature of turbulent environments may impact magnetic reconnection. 
Sec.~\ref{sec:Conclusion} summarizes our current state of knowledge and provides an outlook for future areas of research.

\subsection{The Interaction Between Magnetic Reconnection and Turbulence} \label{sec:TurbulentRecon}
The interaction between magnetic reconnection and turbulence is a complex topic with multiple facets and, as such, it has been examined from a variety of distinct viewpoints in the literature. 
It is, therefore, important to consider what is meant by the interaction between turbulence and reconnection for a given environment, which may include:

\begin{figure*}[h!]
\centering
\includegraphics[width=0.8\textwidth]{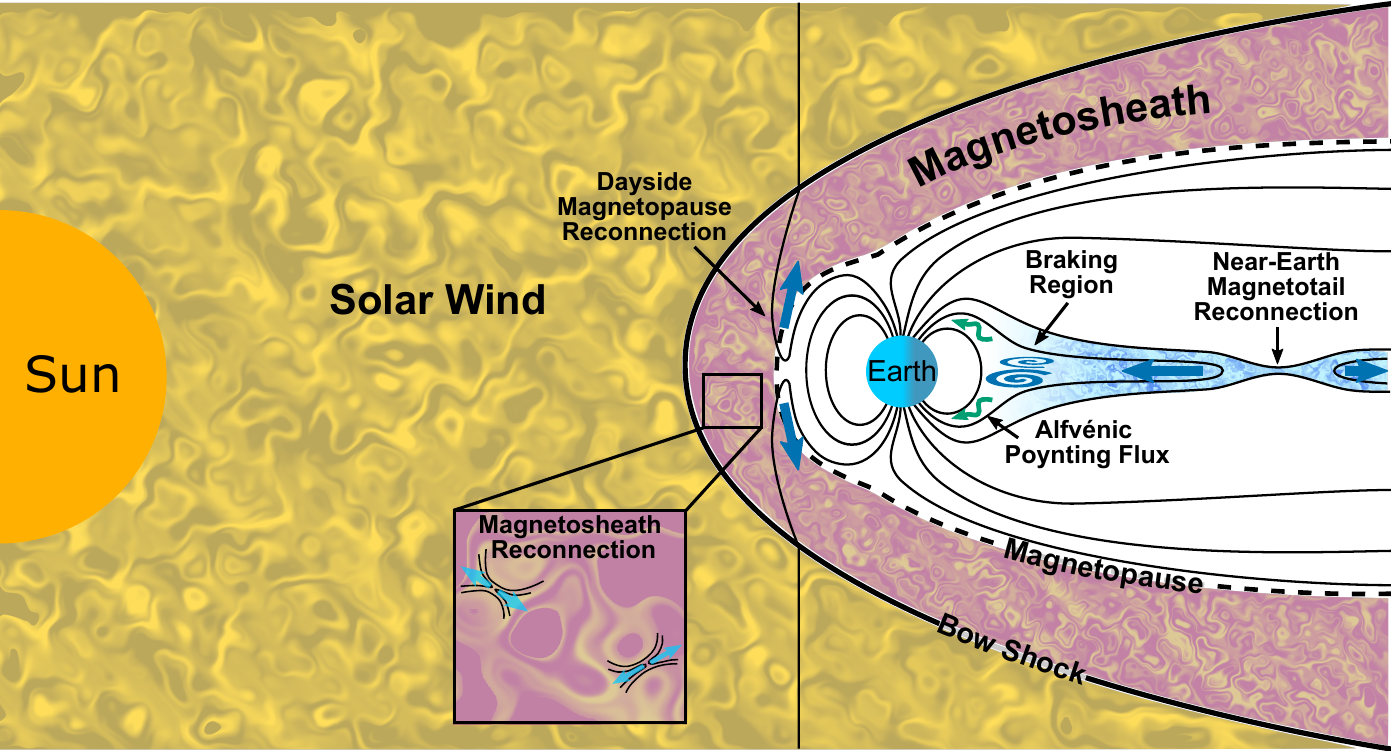}
\caption{Diagram illustrating the turbulent regions in near-Earth space along with the locations of the system-scale reconnection events associated with the Dungey cycle in Earth's magnetosphere. 
Turbulence can both interact with these system-scale reconnection events, as well as generate additional small-scale reconnection events within the turbulent regions.}
\label{fig:turbulent_magnetosphere}
\end{figure*}

\begin{enumerate}
\itemsep=1ex
\item {\it Turbulence-Driven Reconnection} -- Turbulent plasmas are well-known to generate many thin current structures embedded within the fluctuations associated with the sheared, twisted, and tangled magnetic field topologies set up by the turbulent dynamics, which can be sites where reconnection occurs (as illustrated in Fig.~\ref{fig:turbulent_magnetosphere} inset). 
The nature of these current structures is fundamentally linked to an aspect of turbulence referred to as intermittency, in which the turbulent dynamics have a tendency to generate coherent structures with extreme gradients. 
Turbulence-driven reconnection can play a role in facilitating both the nonlinear dynamics and dissipation of the turbulent fluctuations. 
Assessing the importance of these reconnection events in a turbulent medium requires both information about the prevalence of magnetic reconnection at turbulent current sheets and an understanding of the key question of how magnetic reconnection partitions energy as discussed in \citet{Liu2024}.

\item {\it Reconnection-Driven Turbulence} -- Space plasmas also contain regions where reconnection occurs at system-scale current sheets set up by the configuration of the system as a whole, such as the reconnecting current sheets associated with the Dungey cycle at Earth's magnetopause and in the magnetotail (as illustrated in Fig.~\ref{fig:turbulent_magnetosphere}), the heliospheric current sheet, or reconnection associated with the configuration of coronal loops on the Sun.
Large-scale reconnection outflows associated with these current sheets can excite turbulent fluctuations in the system through the spontaneous destabilization of the outflows and the interaction of the outflow with the surrounding plasma. 
In this context, turbulence may impact the reconnection rate ($\mathcal{R}$) through anomalous resistivity/viscosity and can be thought of as a conduit through which energy released by reconnection is re-partitioned across different energy channels in the outflow (although it does not necessarily need to alter the net energy released). 

\item {\it The Transition to Turbulence} -- As well as system-scale reconnection driving turbulence, it can also be driven by other processes. 
Magnetic reconnection can play a key role as a secondary process in the transition of such systems into a fully-developed turbulent state. 
The degree to which this process can be clearly distinguished from {\it turbulence-driven reconnection} and {\it reconnection-driven turbulence} in observations may be more or less clear depending on the system in question; however, a good representative case where this transitory phase is readily accessible to spacecraft observations is the KHI on the flanks of Earth's magnetosphere. 
Numerical simulations of the development of turbulence from other initial configurations, including decaying Alfv\'enic turbulence \citep{Franci2017} and systems of multiple current sheets \citep{Gingell2017}, have also highlighted the potential importance of reconnection in transitioning the initially large-scale fluctuations to a fully-developed turbulence. 

\item {\it Stochastic Reconnection} -- Whether reconnection self-generates turbulence or is embedded in a turbulent environment, the stochastic nature of magnetic field lines in a turbulent flow can potentially have a profound impact on the global structure of the reconnection site. 
For example, the ``rough'' field line topology associated with the broadband distribution of fluctuations may lead to a broad, patchy region where the frozen-in flux theorem is violated and field line wandering in the turbulent flow may disperse field lines faster than otherwise expected. 
Since $\mathcal{R}$ is set by the aspect ratio of the diffusion region, these turbulent effects may significantly increase the reconnection rate if they play a dominant role in the dynamics. 
In this situation, $\mathcal{R}$ would no longer be controlled by microphysical effects, such as resistivity or collisionless processes, but instead by the properties of the turbulent fluctuations, such as the scale at which turbulent energy is injected into the system and fluctuation amplitudes. 
\end{enumerate}
While the above scenarios are not necessarily mutually exclusive, they provide a framework for conceptualizing different aspects of the complex multi-faceted interaction between reconnection and turbulence.

\subsection{Concepts from Turbulence Theory} \label{sec:TurbulentTheory}
While many concepts in turbulence theory can be extended to more complete descriptions of collisionless plasmas, in this section we illustrate some of the basic concepts based on the simplest case of incompressible magnetohydrodynamics (MHD). 
When considering turbulent dynamics it can be useful to divide the particle and electromagnetic variables into spatially and temporally uniform average parts (denoted by subscript $0$) and fluctuations about that mean (denoted by a leading $\delta$), such that for an arbitrary quantity ${\bf g}\left({\bf x},t\right)={\bf g}_0+\delta {\bf g}\left({\bf x},t\right)$. 
Performing this decomposition, the incompressible MHD equations can be written as
\begin{eqnarray}
\frac{\partial \delta{\bf u}}{\partial t} &=& -\overbrace{\frac{\left( \nabla \times \delta{\bf B} \right)\times{\bf B}_0}{\mu_0\rho_0} - \frac{\nabla \delta p}{\rho_0}}^{Linear}
                                                                -\overbrace{\delta{\bf u}\cdot\nabla\delta{\bf u} -\frac{\left( \nabla \times \delta{\bf B} \right)\times\delta{\bf B}}{\mu_0\rho_0}\hspace{37pt}}^{Nonlinear} 
                                                               + \overbrace{\nu \nabla^2\delta{\bf u}}^{Dissipative},   \hspace{15pt} \label{eq:mhd_u} \\
\frac{\partial\delta{\bf B}}{\partial t} &=&    \hspace{10pt} \nabla\times\left( \delta{\bf u} \times {\bf B}_0\right)
                                                               \hspace{36pt} + \nabla\times\left( \delta{\bf u} \times \delta{\bf B}\right)
                                                               \hspace{37pt}\hspace{53pt} + \frac{\eta}{\mu_0} \nabla^2\delta{\bf B}, \label{eq:mhd_b} \\
\nabla^2 \delta p  &=&                            - \hspace{2pt}\nabla \cdot \left(\frac{\left( \nabla \times \delta{\bf B} \right)\times{\bf B}_0}{\mu_0}\right)
                                                               \hspace{2pt}-\nabla \cdot \left(\rho_0 \delta{\bf u}\cdot\nabla\delta{\bf u} +  \frac{\left( \nabla \times \delta{\bf B} \right)\times\delta{\bf B}}{\mu_0}\right),
                                                               \label{eq:mhd_divu} 
\end{eqnarray} 
where ${\bf u}$ is the single-fluid velocity with ${\bf u}_0$ taken to be zero since the equations are Galilean invariant, ${\bf B}$ is the (divergence free) magnetic field, $p$ is the particle pressure, $\rho=\rho_0$ is the mass density that we have taken to be spatially and temporally uniform, and $\mu_0$ is the vacuum permeability. 
We have included collisional viscous and resistive dissipation, which introduce the kinematic viscosity ($\nu$) and the resistivity ($\eta$), for illustrative purposes; however, in collissionless plasmas dissipation occurs through kinetic process that are not present in the single-fluid MHD approximation. 
Eq.~\ref{eq:mhd_divu} follows from the incompressibility condition $\nabla\cdot\delta{\bf u}=0$. 

By dividing variables into mean and fluctuating parts, three classes of terms are apparent -- 
1) linear terms, which in isolation (e.g., for sufficiently small amplitude fluctuations, or if the dynamics preserve alignments that minimise the nonlinearities) give rise to linear wave modes, 
2) nonlinear terms, which give rise to the turbulent dynamics, and 
3) dissipative terms, which remove fluctuation energy from the system. 
In general, both linear and nonlinear terms contribute to the dynamics in the fully nonlinear system and developing a theoretical description of the turbulence requires understanding the interplay between these dynamics. 
While under the right circumstances exact nonlinear solutions for, typically isolated, plasma structures may be possible, fully-developed turbulent systems are characterised by dynamics that are sufficiently nonlinear and made up of such a multitude of interacting structures and fluctuations that exact solutions to the equations become intractable.

The relative importance of the nonlinear to dissipative dynamics for a system can be estimated by comparing the typical amplitude of the nonlinear advection term ($-{\bf u}\cdot\nabla{\bf u}$) to the viscous dissipation term ($\nu\nabla^2{\bf u}$) in Eq.~\ref{eq:mhd_u} such that
\begin{equation}\label{eq:Re}
\frac{\left|-\delta{\bf u}\cdot\nabla\delta{\bf u}\right|}{\left|\nu\nabla^2\delta{\bf u}\right|}\sim\frac{\mathcal{L}\mathcal{U}}{\nu}\equiv Re, 
\end{equation}
where $\mathcal{L}$ and $\mathcal{U}$ represent a characteristic length scale and velocity for the fluctuations. 
For MHD systems, two additional analogues to the Reynolds number can be defined based on the nonlinear Lorentz force term in Eq.~\ref{eq:mhd_u} and the nonlinear magnetic advection term in Eq.~\ref{eq:mhd_b}. 
For sufficiently large $Re$, the nonlinear terms significantly dominate over the dissipative terms across a wide range of scales and the system becomes turbulent. 

Nearly collisionless plasmas are by definition high-$Re$ in the sense that $\nu\sim\eta\sim0$; however, some care must be taken because, in the absence of collisions, kinetic phenomena associated with the breakdown of the fluid approximation can introduce dissipative effects at scales larger than those expected from collisions. 
An alternative way to interpret $Re$ is as a measure of the scale separation between $\mathcal{L}$ and the so-called Kolmogorov microscale, defined as the scale $\ell_{kol}$ at which dissipative dynamics dimensionally become more important than the nonlinear dynamics, such that $Re=(\mathcal{L}/\ell_{kol})^{3/4}$. 
It has been suggested that an alternative way to characterize $Re$ for collisionless plasmas is to consider the scale separation between $\mathcal{L}$ and the ion scales \citep{Parashar2019}.

The relative importance of the nonlinear to linear terms can be estimated by the dimensionless parameter 
\begin{equation}\label{eq:chi}
\chi = \frac{\tau_L}{\tau_{NL}}
\end{equation}
where $\tau_L=\omega_L^{-1}$ is the linear timescale given by the inverse of the frequency of the associated wave and $\tau_{NL}$ is the timescale associated with the nonlinear dynamics of interest. 
If $\chi>1,$ nonlinear interactions are faster than linear dynamics (i.e., the propagation of linear waves) and the nonlinear dynamics dominate the behavior, while, if $\chi<1$, the linear dynamics are faster than the nonlinear interactions.
Eq.~\ref{eq:chi} is dimensionally equivalent to comparing the amplitudes of the nonlinear and linear terms in a given equation in analogy to the definition of $Re$; however, $\chi$ is more typically framed in terms of the ratio of timescales as this definition lends itself to comparing the importance of the nonlinear dynamics to the normal modes of the system. 
While in Eqs.~\ref{eq:mhd_u} -- \ref{eq:mhd_divu}, we have split the linear and nonlinear dynamics based on the global ${\bf B}_0$, it can also be illustrative to treat ${\bf B}_0$ as a scale dependant or locally defined quantity, with the conceptual picture that small-scale fluctuations see their locally averaged field as the background.

\subsubsection{Cross-Scale Energy Transfer}\label{sec:theory_cascade}
Given the complex multiscale nature of turbulent systems, theoretical descriptions are typically formulated in terms of statistical properties of the ensemble of interacting fluctuations - either in physical or spectral space.
In physical space, the autocorrelation tensor for an arbitrary variable ${\bf g}\left({\bf x}\right)$ is given by 
\begin{equation}
R_{ij}^g\left({\bf x},\boldsymbol{\ell}\right) = \langle\delta g_i\left({\bf x} + \boldsymbol{\ell} \right) \delta g_j\left( {\bf x} \right) \rangle,
\end{equation}
where $\langle...\rangle$ represents an ensemble average. 
The correlation function associated with the total fluctuation energy in ${\bf g}\left({\bf x}\right)$ is given by the trace of $R_{ij}^g$, such that $R^g\left({\bf x},\boldsymbol{\ell}\right) = \langle\delta {\bf g}\left({\bf x} + \boldsymbol{\ell} \right)\cdot \delta {\bf g}\left( {\bf x} \right) \rangle$, while the off-diagonal part of the tensor encodes information about the helicity of the fluctuations \citep{Matthaeus1982}. 
One common measure of the characteristic scale associated with $\delta {\bf g}$ is the correlation length ($\lambda_{C,g}$) defined as
\begin{equation}
\lambda_{C,g}= \frac{1}{R^g\left(0\right)}\int_0^\infty{R^g\left(\boldsymbol{\ell}\right)d\ell},
\end{equation}
which represents the scale over which structures in $\delta {\bf g}$ decorrelate and can have different values along different directions if $R^g\left(\boldsymbol{\ell}\right)$ is an anisotropic function.

The $2^{nd}$-order structure function is given by 
\begin{equation}
S_2^{g_i}\left({\bf x},\boldsymbol{\ell}\right) = \langle \left[\Delta g_i\left( {\bf x},\boldsymbol{\ell} \right)\right]^2\rangle = \langle g_i\left({\bf x} + \boldsymbol{\ell} \right) - g_i\left( {\bf x} \right) \rangle.
\end{equation}
with the total fluctuation energy in $\delta{\bf g}$ associated with $S_2^g\left({\bf x},\boldsymbol{\ell}\right) = \langle \left|\Delta {\bf g}\left( {\bf x},\boldsymbol{\ell} \right)\right|^2\rangle$.
A simplifying assumption often, but not always, made in turbulence theory, is that fluctuations are statistically homogeneous and average quantities do not depend on ${\bf x}$. 
In this case there is a relationship between $R_{ii}^g\left(\boldsymbol{\ell}\right)$ and $S_2^{g_i}\left(\boldsymbol{\ell}\right)$, such that
\begin{equation} \label{eq:R-S2}
S_2^{g_i}\left(\boldsymbol{\ell}\right) = 2R_{ii}^g\left(0\right) - 2R_{ii}^g\left(\boldsymbol{\ell}\right).
\end{equation}

The statistical evolution of the turbulence can be considered by re-expressing Eqs.~\ref{eq:mhd_u}-\ref{eq:mhd_divu} as equations for the evolution of the structure functions or correlation functions associated with the total (bulk kinetic + magnetic) fluctuation energy (i.e., $S_2\left(\boldsymbol{\ell}\right) = S_2^u\left(\boldsymbol{\ell}\right) + S_2^{B_A}\left(\boldsymbol{\ell}\right)$ or $R\left(\boldsymbol{\ell}\right) = R^u\left(\boldsymbol{\ell}\right) + R^{B_A}\left(\boldsymbol{\ell}\right)$ where ${\bf B}_A={\bf B}/\sqrt{\mu_0\rho}$).
For $S_2\left(\boldsymbol{\ell}\right)$, this gives \citep{Politano1998b, Adhikari2023}       
 \begin{equation}\label{eq:KH_S2}
\frac{\partial S_2\left(\boldsymbol{\ell}\right)}{\partial t} = \mathcal{P}\left(\boldsymbol{\ell}\right) -\frac{\partial}{\partial \boldsymbol{\ell}}\cdot{\bf Y}\left(\boldsymbol{\ell}\right) + 2D\left(\boldsymbol{\ell}\right) - 4\epsilon,
 \end{equation}
where $\mathcal{P}\left(\boldsymbol{\ell}\right)$ describes the injection of fluctuation energy into the system through some external driving (such driving was not explicitly included in Eqs.~\ref{eq:mhd_u}-\ref{eq:mhd_divu} but is included here to aid in the interpretation of Eq.~\ref{eq:KH_S2}), 
${\bf Y}\left(\boldsymbol{\ell}\right) = \langle \Delta{\bf u}\left(\boldsymbol{\ell}\right)|\Delta{\bf u}\left(\boldsymbol{\ell}\right)|^2 + \Delta{\bf u}\left(\boldsymbol{\ell}\right)|\Delta{\bf B}_A\left(\boldsymbol{\ell}\right)|^2 - 2 \Delta{\bf B}_A\left(\boldsymbol{\ell}\right) \left[\Delta{\bf u}\left(\boldsymbol{\ell}\right)\cdot\Delta{\bf B}_A\left(\boldsymbol{\ell}\right) \right]\rangle$ is a mixed $3^{rd}$-order structure function encoding the effect of the nonlinear terms, 
$D\left(\boldsymbol{\ell}\right) = \nu\frac{\partial^2}{\partial\boldsymbol{\ell}^2}S_2^u\left(\boldsymbol{\ell}\right) + \frac{\eta}{\mu_0}\frac{\partial^2}{\partial\boldsymbol{\ell}^2}S_2^u\left(\boldsymbol{\ell}\right)$ describes the impact of dissipation, and $\epsilon$ is the average energy dissipation rate in the system. Eq.~\ref{eq:KH_S2} highlights the closure problem that is one of the core challenges of developing a complete theory of turbulence -- the evolution of $S_2\left(\boldsymbol{\ell}\right)$ or any statistical quantity requires knowledge of higher-order structure functions such as ${\bf Y}\left(\boldsymbol{\ell}\right)$, the evolution of which, in turn, require knowledge of progressively higher-order structure functions. 

For turbulent systems in a statistically steady state, such that $\partial_t S_2\left(\boldsymbol{\ell}\right)=0$, it is assumed that there will be a significant scale separation between the scales where $\mathcal{P}\left(\boldsymbol{\ell}\right)$ is significant and the scales where $D\left(\boldsymbol{\ell}\right)$ is significant. 
In order for Eq.~\ref{eq:KH_S2} to be satisfied at intermediate scales, there must then be a range of scales - typically referred to as the {\it inertial range} - over which the ${\bf Y}\left(\boldsymbol{\ell}\right)$ term dominates and Eq.~\ref{eq:KH_S2} reduces to
\begin{equation} \label{eq:yaglom}
\frac{\partial}{\partial \boldsymbol{\ell}}\cdot{\bf Y}\left(\boldsymbol{\ell}\right) = -4\epsilon.
\end{equation}
Eq.~\ref{eq:yaglom} gives an exact relationship between ${\bf Y}\left(\boldsymbol{\ell}\right)$ and $\epsilon$, describing the role of the nonlinear dynamics in transporting energy across scales from the driving to the dissipation scales \citep{Politano1998a,Politano1998b} (see also, \citet{Kolmogorov1941b} or \citet{Frisch1995} for a discussion of Eq.~\ref{eq:yaglom} for hydrodynamics; and \citet{Marino2023} for a complete review of the derivation in plasmas). 

Eq.~\ref{eq:yaglom} can also be extended to more general cases than homogeneous incompressible MHD, for example by including background velocity shears \citep{Wan2009, Stawarz2011}, compressibility \citep{Banerjee2013}, or the Hall effect \citep{Hellinger2018,Ferrand2019}. 
The above expression, or variants of it, have been widely applied in both numerical simulations and observations to estimate the energy dissipation rate in turbulent plasmas \citep[e.g.][]{Macbride2005, Marino2008, Stawarz2009, Hadid2018, Bandyopadhyay2020}. 
Applications of Eq.~\ref{eq:yaglom} to spacecraft data typically require assumptions about the average geometry of the fluctuations to simplify the divergence with respect to $\boldsymbol{\ell}$, with common examples being isotropy for which ${\bf Y}\left(\boldsymbol{\ell}\right)$ has no angular dependence \citep{Politano1998b} or hybrid anisotropic geometries with separable variations parallel and perpendicular to the magnetic field \citep{MacBride2008, Stawarz2009}. 
However, multispacecraft measurements can be used to explicitly evaluate the divergence \citep{Osman2011, Pecora2023} and alternative formulations may also provide ways to estimate $\epsilon$ without implicit assumptions about anisotropy \citep{Banerjee2017}.

While the above formalism is typically derived and applied in the context of homogeneous turbulent systems, a series of recent studies \citep{Adhikari2020, Adhikari2021, Adhikari2023, Adhikari2024}, have explored the application of Eq.~\ref{eq:KH_S2} to traditional periodic reconnection simulations (where reconnection is initiated in idealised current sheets at the scale of the periodic box) both with and without guide fields. 
In these works, it was found that the behavior of the terms in Eq.~\ref{eq:KH_S2} are qualitatively similar to that found in fully developed turbulent systems, suggesting that the nonlinear dynamics associated with the reconnection process may, in some sense, embody an energy-cascade-like process. 

Analogs of Eq.~\ref{eq:KH_S2} can also be derived in spectral space \citep[e.g.,][]{Alexakis2005, Grete2017} or using scale-filtering approaches, where a coarse-graining  kernel is used instead of structure functions \citep{Aluie2017, Yang2017, Manzini2022}. In spectral space, the analogue of $R^g\left(\boldsymbol{\ell}\right)$ is the energy spectral density, $\mathcal{E}_g\left({\bf k}\right)$. 
For statistically homogeneous fluctuations, $\mathcal{E}_g\left({\bf k}\right)$ is the Fourier transform of $R^g\left(\boldsymbol{\ell}\right)$ and, given Eq.~\ref{eq:R-S2}, is also the Fourier transform of $S_2^g\left(\boldsymbol{\ell}\right)$ for ${\bf k} \ne 0$. 
By manipulating the Fourier transforms of Eqs.~\ref{eq:mhd_u}-\ref{eq:mhd_divu} into an expression for the evolution of $\mathcal{E}\left({\bf k}\right) = \mathcal{E}_u\left({\bf k}\right) + \mathcal{E}_{B_A}\left({\bf k}\right)$ or taking the Fourtier transform of Eq.~\ref{eq:KH_S2} with respect to $\boldsymbol{\ell}$, a spectral representation of Eq.~\ref{eq:KH_S2} can be obtained \citep[e.g.,][]{Pope2000}
 \begin{equation}\label{eq:KH_spec}
\frac{\partial\mathcal{E}\left({\bf k}\right)}{\partial t} = \mathcal{P}\left({\bf k}\right) + \mathcal{T}\left({\bf k}\right) - 2D\left({\bf k}\right) \end{equation}
where $\mathcal{T}\left({\bf k}\right)$ is the analogue of $-\frac{\partial}{\partial \boldsymbol{\ell}}\cdot{\bf Y}\left(\boldsymbol{\ell}\right)$ -- referred to as the transfer function in this context -- and $\mathcal{P}\left({\bf k}\right)$ and $\mathcal{D}\left({\bf k}\right)$ are the Fourier transforms of $\mathcal{P}\left(\boldsymbol{\ell}\right)$ and $\mathcal{D}\left(\boldsymbol{\ell}\right)$, respectively. 
$\mathcal{T}\left({\bf k}\right)$ in general takes the form of a complex set of convolutions associated with the nonlinear terms, representing the net transfer of energy to/from wavevector ${\bf k}$ due to the sum of all possible interactions between other wavevectors and can be related to the cross-scale energy flux, $\boldsymbol{\Pi}\left({\bf k}\right)$, through a surface in ${\bf k}$-space, such that $\mathcal{T}\left({\bf k}\right)=-\frac{\partial}{\partial {\bf k}} \cdot\boldsymbol{\Pi}\left({\bf k}\right)$. 
Often Eq.~\ref{eq:KH_spec} is integrated over spherical shells and $\Pi\left(k\right)$ is interpreted as the isotropic energy flux; however, other surfaces may be relevant to different types of anisotropy \citep{Yokoyama2021}.

The spectral representation highlights two key assumptions often invoked in turbulence theory - 1) that the nonlinear interactions are local in $k$-space, such that the dominant contributions to the convolutions come from wave vectors with similar magnitudes and 2) that the magnitude of $\Pi\left(k\right)=\epsilon$ and is constant as a function of $k$.
With these two assumptions, the conceptual picture of the {\it energy cascade} through the internal range emerges, where nonlinear interactions incrementally transport energy from scale-to-scale at a constant rate, which in a statistically steady-state, is equal to the average energy dissipation rate. 
Consequently, $\epsilon$ is often referred to as the energy cascade rate in turbulent systems. 
While the large-scale nonlinear dynamics set the flux of energy to the small scales, for collisionless plasmas understanding which processes are responsible for dissipating that energy remains a key challenge and in Sec.~\ref{sec:TurbulenceDrivenReconnection_impacts} we discuss the potential role reconnection may play in turbulent dissipation.

\subsubsection{Energy Spectra} \label{sec:theory_spectra}
The shape of $\mathcal{E}\left( {\bf k}\right)$ or $S_2\left(\boldsymbol{\ell}\right)$ in the inertial range can be estimated by making assumptions about the nature of the dominant nonlinear interactions.
The basic ingredients for estimating these scalings amount to 1) taking $\epsilon$ to be independent of scale within the inertial rage and 2) estimating the timescale $\tau_{tr}$ over which nonlinear interactions transfer energy between scales under a given set of assumptions. 
Relating $\mathcal{E}\left( {\bf k}\right)$ to $\tau_{tr}$ dimensionally gives
\begin{equation} \label{eq:spec_tau} 
 \mathcal{E}\left(k\right) \sim \epsilon\frac{\tau_{tr}}{k},  
\end{equation}
where, for simplicity, we have assumed the fluctuations are isotropic (although this assumption can be relaxed).\footnote{The units of $\mathcal{E}\left( {\bf k}\right)$ are such that when integrated over all ${\bf k}$ it gives the average total fluctuation energy in the system, such that the quantity $\mathcal{E}\left( {\bf k}\right)d^3{\bf k}$ has units of energy. In some situations it is useful to consider the isotropic (omnidirectional) energy spectrum, defined such that $\mathcal{E}\left(k\right)dk$ has units of energy, and in plasmas it is often useful to consider the gyrotropically integrated spectrum, defined such that $\mathcal{E}\left(k_{||}, k_{\perp}\right)dk_{||}dk_{\perp}$ has units of energy. Furthermore, in the gyrotropic case, it is sometimes useful to consider the spectra marginalized over $k_{||}$ or $k_{\perp}$, such that $\mathcal{E}\left(k_{\perp}\right)dk_{\perp}$ and $\mathcal{E}\left(k_{||}\right)dk_{||}$ have units of energy, respectively.} 
In principle, $\tau_{tr}$ can be a function of scale, the geometry of the fluctuations, and the physical properties of the medium, and its dependence on these parameters is intrinsically linked to the nature of the underlaying nonlinear interactions enabling the cascade, leading to different predictions for $\mathcal{E}\left( {\bf k}\right)$.
This link between predictions for the shape of $\mathcal{E}\left( {\bf k}\right)$ and the nature of the nonlinear dynamics is one reason why $\mathcal{E}\left( {\bf k}\right)$ is an important observable quantity for understanding turbulent dynamics, although it does not provide the full picture as discussed in Sec.~\ref{sec:theory_intermittency}. 
Single spacecraft measurements are capable of providing spacecraft-frame frequency spectra; however, assuming the background flow (${\bf U}_0$) is sufficiently fast, comparisons can be made between observed spectra and theoretical predictions for $\mathcal{E}\left( {\bf k}\right)$ by employing the so-called Taylor hypothesis discussed in Appendix~\ref{app:taylor}. 
Table~\ref{tbl:EnergySpec} illustrates some example models for $\mathcal{E}\left( {\bf k}\right)$. 
The first five models are based on MHD, while the final model gives an example for sub-proton-scale dynamics. 

\begin{table}
\caption{Summary of example models for the turbulent energy spectrum.}
\label{tbl:EnergySpec}     
\begin{tabular}{l@{\hskip8pt}c@{\hskip8pt} r@{\hskip3pt}c@{\hskip3pt}l @{\hskip8pt}c}
\hline\noalign{\smallskip}
Model 			& $\tau_{tr}$ 			& \multicolumn{3}{c}{Spectral Prediction} 			& Anisotropy \\
\noalign{\smallskip}\hline\noalign{\smallskip}
Kolmogorov$^{a}$						& $\tau_{NL}\sim\frac{1}{k\delta u\left(k\right)}$																										& $\mathcal{E}\left(k\right)$ & $\sim$ & $\epsilon^{2/3} k^{-5/3}$ 																						& Isotropic																					\\
Iroshnikov$^{b}$-Kraichnan$^{c}$ 			& $\frac{\tau_{NL}}{\chi}\sim\frac{V_A}{k\left[\delta u\left(k\right)\right]^2}$																					& $\mathcal{E}\left(k\right)$ & $\sim$ & $\left(\epsilon V_A\right)^{1/2} k^{-3/2}$ 																			& Isotropic																					\\
Weak Alfv\'enic$^{d}$					& Derived exactly assuming $\chi\ll1$		 																									& $\mathcal{E}\left(k_{||},k_{\perp}\right)$ & $\sim$ & $f\left(k_{||}\right)\left(\epsilon V_A\right)^{1/2} k_{\perp}^{-2}$ 													& Purely $\perp$																				\\
Goldreich-Sridhar$^{e}$					& $\chi\sim1 \rightarrow \frac{1}{k_\perp\delta u_\perp\left(k_\perp\right)}\sim\frac{1}{k_{||}V_A}$																& $\mathcal{E}\left(k_{\perp}\right)$ & $\sim$ & $\epsilon^{2/3}k_\perp^{-5/3}$ 																				& $k_{||}\sim\epsilon^{1/3}V_A^{-1}k_{\perp}^{2/3}$ 														\\
Dynamic Alignment$^{f}$ 				& $\chi\sim1 \rightarrow \frac{V_A}{k_\perp\left[\delta u_\perp\left(k_\perp\right)\right]^2}\sim\frac{1}{k_{||}V_A}$														& $\mathcal{E}\left(k_{\perp}\right)$ & $\sim$  & $\left(\epsilon V_A\right)^{1/2} k_{\perp}^{-3/2}$ 																& $k_{||}\sim\epsilon^{1/2}V_A^{-3/2}k_{\perp}^{1/2}$ 													\\
Strong Hall MHD$^{g}$ 			& $\tau_{NL}\sim\frac{1}{d_ik^2\delta B\left(k\right)}$																										& $\mathcal{E}\left(k\right) \sim \mathcal{E}_B\left(k\right)$ & $\sim$ & $\epsilon^{2/3}  d_i^{-2/3} k^{-7/3}$ 														& Isotropic 																					\\
\noalign{\smallskip}\hline
\multicolumn{6}{l}{$^{a}$\footnotesize{\citet{Kolmogorov1941a};} $^{b}$\footnotesize{\citet{Iroshnikov1964};} $^{c}$\footnotesize{\citet{Kraichnan1965};} $^{d}$\footnotesize{\citet{Galtier2000};}}\\
\multicolumn{6}{l}{$^{e}$\footnotesize{\citet{Goldreich1995};} $^{f}$\footnotesize{\citet{Boldyrev2006};} $^{g}$\footnotesize{\citet{Biskamp1999}}}
\end{tabular}
\end{table}

Strongly nonlinear models where $Re\gg1$  and $\chi\gg1$, as in the {\it Kolmogorov} model, assume $\tau_{tr}$ is governed by the nonlinear timescale ($\tau_{NL}$) associated with the advection terms.
\citet{Kolmogorov1941a} originally applied this model to incompressible hydrodynamic turbulence, predicting the well-known $k^{-5/3}$-spectrum. 
However, since the additional nonlinear advective and Lorentz force terms in the MHD equations are dimensionally equivalent to the advective term in hydrodynamics, an analogous approach can be taken with incompressible MHD, producing an equivalent spectral prediction \citep{Biskamp2000}. 

On the other hand, the introduction of a magnetic field, as well as other effects such as compressibility, collisionless dynamics, etc., also introduces linear terms describing the effect of waves. 
For $Re\gg1$ and $\chi\ll1$, often referred to as ``weak'' or ``wave'' turbulence, nonlinear interactions are strongly mediated by wave-like dynamics. 
The {\it Iroshnikov-Kraichnan} model is a simple isotropic model for incompressible MHD in this regime, whereby $\tau_{tr}$ is lengthened by a factor of $\chi^{-1}$, since multiple ``collisions'' between propagating Alfv\'en wave packets are required for the nonlinearities to fully distort the wave packets and transfer fluctuation energy across scales if the wave propagation is much faster than the nonlinear timescale, resulting in a $k^{-3/2}$-spectrum.
However, in the weak turbulence regime, exact analytical progress can also be made \citep{Nazarenko2011, Galtier2023}.
\citet{Galtier2000} applied weak turbulence formalism to derive a {\it weak Alfv\'enic} model for incompressible MHD, illustrating that, in the limit of weak turbulence, the cascade is fundamentally anisotropic with the cascade of energy proceeding purely in the direction perpendicular to ${\bf B}_0$ and resulting in a spectrum scaling as $k_\perp^{-2}$ and following an arbitrary function of $k_{||}$ set by the driving. 

Anisotropic strong turbulence models, such as the {\it Goldreich-Sridhar} and {\it Dynamic Alignment} models, often incorporate a constraint known as critical balance, whereby $\chi\sim1$ such that the linear and nonlinear terms balance scale-by-scale. 
In these models, $\tau_{tr}$ is again related to $\tau_{NL}$ associated with nonlinear advection, except now the nonlinear interaction is assumed to be inherently anisotropic with only $k_\perp$ contributing to the nonlinear interaction. 
Since the dispersion relation for Alfv\'en waves is also anisotropic, the critical balance constraint then provides a prediction for the anisotropy of the spectrum. 
While the {\it Goldreich-Sridhar} model essentially applies analogous phenomenology as the {\it Kolmogorov} model to predict a $k_\perp^{-5/3}$-spectrum, the {\it Dynamic Alignment} model argues that geometrical constraints associated with the strength of ${\bf B}_0$ reduce the the efficiency of nonlinear interactions, producing a $k_\perp^{-3/2}$-spectrum reminiscent of the {\it Iroshnikov-Kraichnan} model but based on different phenomenological arguments. 

The same general framework can be used in systems where other nonlinearities play a dominant role, such as at sub-proton-scales where the single-fluid MHD approximation breaks down. 
The {\it Strong Hall MHD} model provides an example of analysing the nonlinear Hall term in the induction equation under the assumption that the Hall term significantly dominates the dynamics (often referred to as electron-MHD) in a manner similar to the {\it Kolmogorov} model. 
In this situation, ${\bf B}$ carries the majority of the fluctuation energy and, because of the presence of an additional derivative in the Hall term, a steeper $k^{-7/3}$-spectrum is obtained. 
However, other analyses of the sub-proton scale dynamics, such as critical balance models invoking kinetic-scale wave modes like whistler waves \citep{Narita2010, Boldyrev2013, Narita2016}, kinetic Alfv\'en waves \citep{Boldyrev2012}, and inertial kinetic Alfv\'en waves \citep{Chen2017}, are also commonly employed to explain spacecraft observations. 

The models discussed above implicitly assume turbulent fluctuations are space-filling and self-similar. 
Violations of this assumption, discussed further in Sec.~\ref{sec:theory_intermittency}, may also have an impact on $\mathcal{E}\left( {\bf k}\right)$. 
Such corrections to predicted power laws have been proposed and are often invoked, particularly at sub-proton-scales, to explain steeper power laws of $\sim k^{-2.8}$ or $\sim k^{-3}$ in space plasmas \citep{Boldyrev2012}. 
Other complexities may also be present that are not illustrated in the examples provided in Table~\ref{tbl:EnergySpec}, such as constraints on the alignments between fluctuations in different vector fields that may be imposed by other conserved quantities (e.g., magnetic helicity, cross helicity, or generalized helicity) \citep{Pouquet2019, Meyrand2021, Squire2022}. 
In Sec.~\ref{sec:TurbulenceDrivenReconnection_impacts}, we discuss further how reconnection, in particular, may alter $\mathcal{E}\left( {\bf k}\right)$. 

\subsubsection{Intermittency \& Current Sheets} \label{sec:theory_intermittency}
The energy spectrum on its own does not provide the full picture of turbulence. 
Notably, turbulent systems are not simply comprised of uncorrelated randomly superposed normal modes and, in fact, phase correlations between modes at different scales are spontaneously generated by the nonlinear dynamics. 
These phase correlations manifest as localized {\it coherent structures} in physical space. 
Coherent structures can take the form of vorticity sheets and filaments in hydrodynamics \citep{Okamoto2007} and, with the inclusion of {\bf B}, current sheets and magnetic discontinuities, among other complex structures \citep{Mininni2006, Greco2009, Uritsky2010, Matthaeus2015}. 
This feature of turbulence is referred to as {\it intermittency}, so-called because it results in a statistically non-uniform ``intermittent" distribution of dissipative structures in the domain. 
A key aspect of intermittency is that it violates the assumed self-similar and space-filling nature of the nonlinearly interacting fluctuations that are an inherent feature of many of the theoretical models discussed in Sec.~\ref{sec:theory_spectra}. 
Magnetic shears associated with coherent structures (e.g. current sheets or more generalized current structures) are potential sites where reconnection can occur in turbulent plasmas. 
The intrinsic link between the intermittent nature of the turbulence and the statistical properties (e.g., structure, prevalence, distribution throughout the domain, etc.) of coherent structures means intermittency is likely an important feature to consider in the case of turbulence-driven reconnection. 

Intermittency is typically analyzed by examining higher-order statistics, such as the $p^{th}$-order structure functions given by
\begin{equation}
S^g_p\left(\boldsymbol{\ell}\right)=\langle\left[\Delta g\left(\boldsymbol{\ell}\right) \right]^p\rangle. 
\end{equation}
While $S_2\left(\boldsymbol{\ell}\right)$ has a direct relationship to $\mathcal{E}\left({\bf k}\right)$, higher-order statistics encode additional information about cross-scale correlations. 
The \citet{Kolmogorov1941a} theory of turbulence, which does not include the effect of intermittency, predicts $S^{u_\ell}_p\left(\boldsymbol{\ell}\right)\sim\ell^{\zeta_p}$ with $\zeta_p=p/3$ and ${u_\ell}$ the component of ${\bf u}$ along $\boldsymbol{\ell}$. 
Turbulent systems typically exhibit strong deviations from this scaling, with increasingly sub-$p/3$ scalings as $p$ increases, in both hydrodynamic \citep{Anselmet1984} and plasma systems \citep{Marsch1997, Biskamp2000}. 
Toy models, such as so-called $\beta$, multi-fractal, and random cascade models (see \citet{Frisch1995} for a detailed discussion), which relax assumptions about self-similarity and the statistical homogeneity of dissipation in various ways, demonstrate that intermittency offers an explanation for this behavior.  

While some numerical studies indicate intermittency is more intense in MHD than hydrodynamics \citep{Biskamp2000}, observations from the solar wind suggest, at sub-proton-scales $\zeta_p$ is linear with $p$, as in the non-intermittent case \citep{Kiyani2009}.
The behavior at sub-proton-scales may not be universal, however, with observations in Earth's magnetosheath \citep{Chhiber2018} and numerical simulations \citep{Franci2015}, continuing to show signatures of intermittency well into the sub-proton-scales. 
A complete understanding of this variation behaviour, remains an open question and may suggest a dynamically significant variation in the nature or distribution of nonlinear fluctuations at sub-proton-scales across different turbulent environments. 

Since $S^g_p\left(\boldsymbol{\ell}\right)$ are the statistical moments of the distribution of increments $\Delta g\left(\boldsymbol{\ell}\right)$, characterizing intermittency can be framed as considering how the probability distribution function of increments varies with scale. 
Typically distributions of $\Delta{\bf u}\left(\boldsymbol{\ell}\right)$ or $\Delta{\bf B}\left(\boldsymbol{\ell}\right)$ exhibit nearly Gaussian shapes for large $\ell$ and become progressively more heavy-tailed (larger than Gaussian probability of extreme values) for small $\ell$ \citep{Frisch1995, Sorriso-Valvo1999}. 
Given the heavy-tailed nature of the distributions, one common measure of intermittency is the scale-dependant kurtosis given by $\kappa\left(\boldsymbol{\ell}\right)=S^g_4\left(\boldsymbol{\ell}\right)/\left[S^g_2\left(\boldsymbol{\ell}\right)\right]^2$, which is expected to have a value of 3 at large $\ell$ where the distribution is Gaussian and then become increasingly larger through the inertial range \citep{Wu2013}. 
Heuristically, this behavior can be understood from the fact that at large separations ($\gg\lambda_{C,g}$), quantities at two different points will be uncorrelated and $\Delta g\left(\boldsymbol{\ell}\right)$ will be an uncorrelated random variable, while at small separations $\Delta g\left(\boldsymbol{\ell}\right)$ will be sensitive to the gradients in these quantities, which are spatially inhomogeneous and intermittent. 
In fact, detailed examinations of the distributions of the vorticity and current density in turbulent plasmas present a picture where particularly intense vorticity and current structures form at the interfaces between large regions of reduced nonlinear activity, consistent with the heavy-tailed distributions \citep{Servidio2008, Greco2009, Pecora2021, Pecora2023}. 
Furthermore, locating extreme values in $\Delta g\left(\boldsymbol{\ell}\right)$ has been proposed as a means of identifying coherent (potentially dissipative) structures, e.g., by examining the so-called partial variance of increments ($PVI=\Delta g\left(\boldsymbol{\ell}\right)/\sqrt{S_2^g\left(\boldsymbol{\ell}\right)}$; \citet{Greco2009}) or the integrands of ${\bf Y}\left(\boldsymbol{\ell}\right)$ (i.e., $\Delta{\bf u}\left(\boldsymbol{\ell}\right)|\Delta{\bf u}\left(\boldsymbol{\ell}\right)|^2$, $\Delta{\bf u}\left(\boldsymbol{\ell}\right)|\Delta{\bf B}_A\left(\boldsymbol{\ell}\right)|^2$, and $\Delta{\bf B}_A\left(\boldsymbol{\ell}\right) \left[\Delta{\bf u}\left(\boldsymbol{\ell}\right)\cdot\Delta{\bf B}_A\left(\boldsymbol{\ell}\right) \right]$; \citet{Sorriso-Valvo2018}).

\section{Turbulence-Driven Reconnection} \label{sec:TurbulenceDrivenReconnection} 
{\it In situ} spacecraft observations of turbulent plasmas in near-Earth space, such as the solar wind \citep{Greco2009}, magnetosheath \citep{Gingell2021, Schwartz2021}, and plasma sheet \citep{Ergun2018}, are filled with current structures and associated magnetic shears that can be sites of reconnection. 
The extent to which reconnection is self-consistently initiated at these current sheets and the impact that this has on the turbulent dynamics has long been a topic of interest and there is a rich body of literature attempting to explore this issue both observationally and in numerical simulations. 

The solar wind is one region of turbulence that has been extensively studied with {\it in situ} spacecraft observations. 
Numerous studies have provided evidence for reconnection exhausts at solar wind current sheets \citep{Gosling2005, Gosling2007, Phan2020, Eriksson2022, Fargette2023}.
However, solar wind reconnection exhausts are often encountered hundreds or even thousands of ion inertial lengths ($d_i$) away from the x-line \citep{Mistry2015b}, suggesting that reconnecting current sheets in the solar wind extend over large length-scales. 
It is, therefore, challenging to distinguish reconnecting current sheets in the solar wind that may be self-consistently generated by the turbulent dynamics from those associated with the evolution of large-scale solar wind structure (e.g., heliospheric current sheet, stream interaction regions, coronal mass ejections) and it remains an open question as to the extent to which these two populations contribute to the identified reconnection exhausts in the solar wind \citep{Eriksson2022}. 
Recently, however, there has been new progress in observationally examining turbulence-driven magnetic reconnection in Earth's magnetosheath, where high-resolution measurements from {\it MMS} have provided evidence for small-scale reconnection events embedded within the recently excited turbulent fluctuations downstream of Earth's bow shock.

\subsection{Bow Shock \& Magnetosheath Reconnection} \label{sec:TurbulenceDrivenReconnection_obs} 
Earth’s bow shock forms at the interface between the solar wind and Earth’s magnetosphere, where the super-Alfv\'enic solar wind suddenly slows down to a sub-Alfv\'enic speed due to Earth's strong magnetic field and the kinetic energy in the solar wind is converted to magnetic, thermal, and fluctuation energy. 
In the shock, the electron motion is frozen-in to ${\bf B}$, while the ion motion is decoupled from the electron motion and ${\bf B}$. 
Ions can penetrate deep inside the shock transition layer without gyration and, as a result, a shock potential is generated, which reflects some ions to produce ion beams propagating upstream of the shock. 
The ion-ion beam instability caused by the reflected and solar wind ion populations generates large amplitude electromagnetic waves and the plasma in the shock becomes turbulent. 
Particles crossing the shock are rapidly heated, forming a region downstream of the shock where $|{\bf B}|$, number density ($n$), and temperature ($T$) are enhanced compared to the upstream solar wind - referred to as the magnetosheath. 

Fluctuations in the magnetosheath have features consistent with turbulent dynamics, including broadband power law spectra as discussed in Sec.~\ref{sec:theory_spectra} \citep{Sahraoui2006, Alexandrova2008, Huang2014, Huang2017}, higher-order statistics consistent with intermittency as discussed in Sec.~\ref{sec:theory_intermittency} \citep{Yordanova2008, Chhiber2018}, and evidence of an active cross-scale energy cascade as discussed in Sec.~\ref{sec:theory_cascade} \citep{Hadid2018, Bandyopadhyay2018, Bandyopadhyay2020}. 
In contrast to the solar wind, magnetosheath fluctuations typically have a much shorter $\lambda_C$ that varies from tens to hundreds of $d_i$ \citep{Stawarz2022}. 
Furthermore, other fluctuation properties - such as the MHD-scale spectral index and turbulent Mach number (ratio of velocity fluctuation amplitude to sound speed)  - also vary across the magnetosheath \citep{Huang2017, Li2020}.
These features suggest that the processing of the solar wind plasma by the shock, potentially through processes such as large-amplitude wave-generation associated with the quasi-parallel shock, temperature anisotropy instabilities downstream of the shock, or the interaction of turbulent fluctuations in the solar wind with the shock \citep{Omidi1994, Bessho2020, Trotta2023}, drives new fluctuations into the system that interact nonlinearly and evolve into a turbulent state.   

Spacecraft observations demonstrate these fluctuations are associated with many current sheets. 
Using {\it Cluster} observations, \citet{retino2007} found that, among the turbulent current sheets in Earth’s magnetosheath, there are current sheets with signatures of reconnection. 
\citet{retino2007} showed that the thickness of one such reconnecting current sheet was $\sim d_i$ and observed the out-of-plane reconnection ${\bf E}$, the quadrupolar Hall ${\bf B}$, the bipolar Hall ${\bf E}$ pointing toward the center of the current sheet, and a positive value of $\mbox{\boldmath$j$}\cdot\mbox{\boldmath$E$}$, indicating energy exchange from the electromagnetic fields to the particles in accordance with Poynting's theorem. 
The reconnection outflow inferred from the ${\bf E}\times {\bf B}$ drift was 0.1 times the Alfv\'en speed, indicating $\mathcal{R}\sim0.1$. 
Notably, due to the small length scale, and associated short time scale over which the event was advected over the spacecraft, all of the signatures of turbulence-driven magnetosheath reconnection identified with {\it Cluster} were obtained from the electromagnetic field measurements. 

Subsequent {\it MMS} observations also revealed many reconnecting current sheets in the magnetosheath. 
\citet{yordanova2016} and \citet{voros2016} identified several reconnecting current sheets in Earth's magnetosheath, where reconnection outflows and enhancements of $\mbox{\boldmath$j$}\cdot\mbox{\boldmath$E$}$ were observed. 
\citet{voros2016} detected electron diffusion region (EDR) signatures by investigating the decoupling of the electron velocity from the ${\bf E}\times{\bf B}$ drift, the agyrotropy parameter $Q$ \citep{Swisdak2016}, and electron distribution functions. 
Inside the EDR, $\mbox{\boldmath$j$}\cdot\mbox{\boldmath$E'$}$, where $\mbox{\boldmath$E'$}=\mbox{\boldmath$E$}+\mbox{\boldmath$u_e$}\times\mbox{\boldmath$B$}$, was positive, indicating conversion of magnetic energy to particle kinetic and thermal energy via the non-ideal ${\bf E}$. 

\begin{figure}
\begin{centering}
\includegraphics[width=0.8\textwidth]{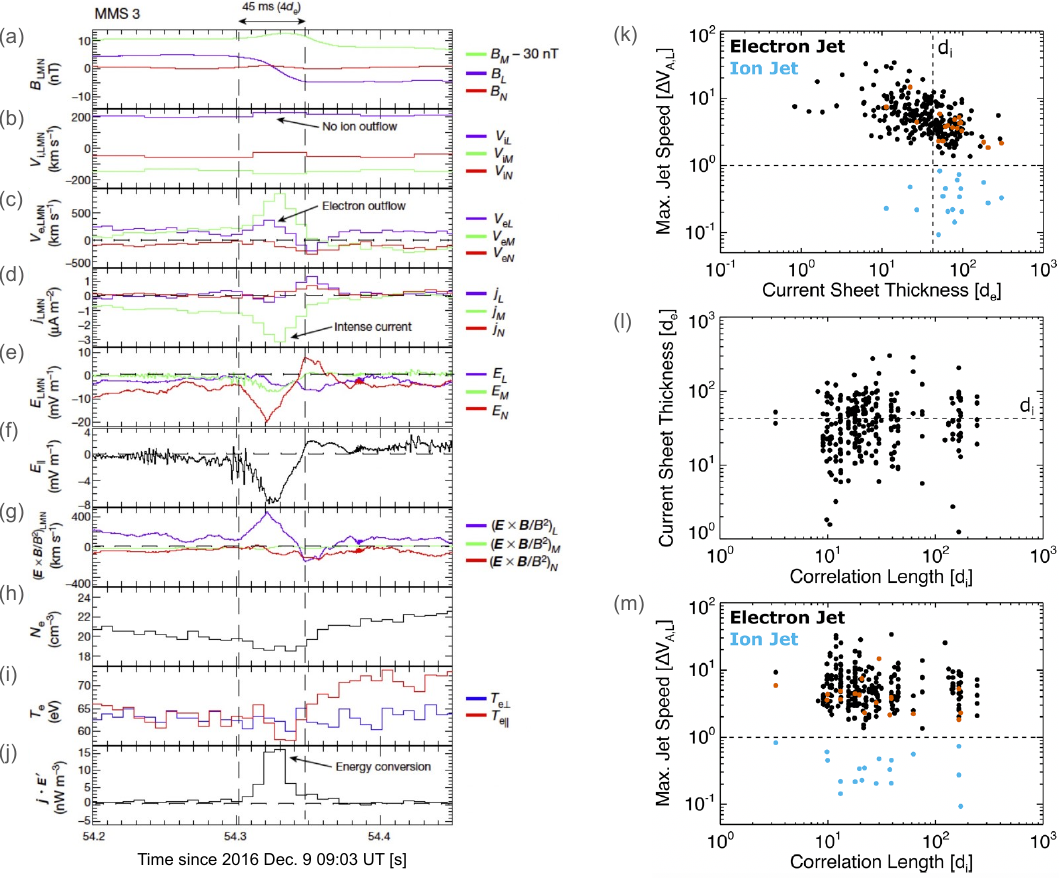}
\caption{Example of electron-only reconnection in the magnetosheath showing (a) ${\bf B}$, (b) ion velocity, (c) electron velocity, (d) ${\bf j}$, (e) ${\bf E}$, (f) the component of ${\bf E}$ along ${\bf B}$, (g) ${\bf E}\times{\bf B}$-drift velocity, (h) electron number density, (i) electron parallel and perpendicular temperatures, and (j) ${\bf j}\cdot{\bf E}^\prime$. All vector quantities are in a local current sheet coordinate system based on a hybrid minimum variance analysis \citep[reproduced from][]{phan2018}. (k)-(m) Statistical study of 256 reconnection events in the turbulent magnetosheath, comparing electron (and ion if present) reconnection jet speeds, current sheet thickness, and $\lambda_{C,{\bf B}}$ \citep[reproduced from][]{Stawarz2022}.}
\label{fig:SheathRec1}
\end{centering}
\end{figure}

\citet{phan2018} examined two nearby segments of high-resolution burst data from {\it MMS} in detail, totalling $\sim21$ minutes of magnetosheath observations, revealing that while many current sheets had evidence of reconnection in the form of electron outflows, none had clear evidence of ion outflows. 
The relative lack of clear ion outflows was despite the fact that fully accelerated ion outflows should occupy a larger volume of space -- and thus should be statistically more likely to be encountered -- than the ion diffusion region (IDR), where fully accelerated ion outflows would not be expected.
Since electrons appeared to be the only species participating in the reconnection process, this type of reconnection has come to be known as ``electron-only" reconnection. 
An example of one electron-only reconnection event identified by \citet{phan2018} is shown in Fig.~\ref{fig:SheathRec1}a--j. 
For the event in Fig.~\ref{fig:SheathRec1}a--j the thickness of the current sheet was significantly thinner than the ion scales at $\sim4$ electron inertial lengths ($d_e$). 
During the current sheet crossing, {\it MMS3} observed super-Alfv\'enic electron outflows of $\sim250$ km/s in the outflow direction (L direction) relative to an Alfv\'en speed associated with the reconnecting component of the field of $V_{A,L}\sim25$ km/s. 
The other three {\it MMS} spacecraft also observed super-Alfv\'enic electron outflows, but in the opposite direction, providing direct evidence for the two oppositely directed jets. 
The electrons in the current sheet were not frozen-in to ${\bf B}$, and a large $E_{||}$ was observed, producing a large ${\bf j}\cdot{\bf E}^\prime$. 
It was proposed that the reason ions did not appear to be participating in the reconnection process was because the length of the current sheets along the outflow direction was short enough that there was not enough time/space for the reconnected ${\bf B}$ to accelerate ion jets before the field fully relaxed. 
This picture was supported by idealized numerical experiments, where the lengths of reconnecting current sheets were artificially constrained -- resulting in reconnection with unique properties compared to ion-coupled reconnection when the length of the current sheet along the outflow direction was $\lesssim 10d_i$ \citep{Pyakurel2019}. 
These features included the reconnection event being embedded within a thin electron-scale current sheet (in contrast to ion-coupled reconnection, where the electron-scale gradients associated with the EDR are embedded in a broader ion scale current sheet), only featuring fast electron outflows, and having higher reconnection rates. 
Similar results were also found in the fluctuations self-consistently generated in shock simulations \citep{Bessho2019, Bessho2020, Bessho2022, Bessho2023}. 

Since the \citet{phan2018} work, examples of both standard ion-coupled reconnection and electron-only reconnection have been observed in Earth's magnetosheath \citep{wilder2018, wilder2022, voros2019, stawarz2019, Stawarz2022}. 
\citet{voros2019} revealed whistler and lower hybrid waves in the reconnecting current sheets. 
\citet{wilder2018, wilder2022} investigated the energy conversion, demonstrating that reconnection with a small guide field (less than 30\% of the reconnecting magnetic field) exhibits ${\bf j}\cdot{\bf E}^\prime$ primarily associated with ${\bf E}_\perp$ and agyrotropic electron distributions. 
In contrast, large guide fields are associated with ${\bf j}\cdot{\bf E}^\prime$ generated by $E_{||}$. 
\citet{Bandyopadhyay2021} examined the so-called pressure-strain interaction terms, quantifying local energy exchange between the bulk flow and internal energy (defined as the second moment of the distribution function), demonstrating magnetosheath reconnection events are sites of enhanced pressure-strain interaction alongside ${\bf j}\cdot{\bf E}$. 
In the small selection of events considered, \citet{Bandyopadhyay2021} found both positive and negative pressure-strain signatures, suggesting both local ``heating'' and ``cooling'' (although with a slight preference for positive ``heating'' signatures), with typically more intense signatures in the electrons compared to the ions. 

\citet{Stawarz2022} statistically examined the reconnecting current sheets in the turbulent magnetosheath with {\it MMS} data. 
A total of 256 reconnection events (including 18 ion-coupled reconnection events) were investigated to understand the relationship between $\lambda_{C,{\bf B}}$ and the type of reconnection present. 
\citet{Stawarz2022} found thinner current sheets tended to have higher outflow speeds (Fig.~\ref{fig:SheathRec1}k), such that when the thickness is $\lesssim d_i$, the electron jet speed becomes super-Alfv\'enic. 
These super-Alfv\'enic flows can reach the order of the electron Alfv\'en speed, indicating reconnection in the electron-only regime. 
\citet{Stawarz2022} further showed that intervals with $\lambda_{C,{\bf B}} \lesssim 20d_i$ are associated with thinner reconnecting current sheets and faster super-Alfv\'enic electron jets on average (Figs.~\ref{fig:SheathRec1}l,m).
The relationship with $\lambda_{C,{\bf B}}$ suggests a scenario where current sheets form at the interface of $\lambda_{C,{\bf B}}$-scale magnetic structures -- implying $\lambda_{C,{\bf B}}$ controls the average length of current sheets along the outflow direction. 
When $\lambda_{C,{\bf B}}$ approaches tens of $d_i$, it becomes more likely there will be insufficient space for the reconnected field lines to accelerate ion outflows, increasing the prevalence of electron-only reconnection. 

In addition to Earth’s magnetosheath, the shock transition region and foreshock regions are often turbulent, and many reconnection events have been detected by {\it MMS} in these regions. 
\citet{wang2019} and \citet{gingell2019} found evidence of electron-only reconnection in the shock transition region. 
\citet{gingell2019} observed that $T_e$ rose from 20eV to 33eV across the shock ramp and an additional 7eV increase occurred in the shock transition region, suggesting 35\% of the total electron heating across the shock occurs in the transition layer in association with electron-only reconnection. 
\citet{wang2019} additionally observed ion-coupled reconnection with both ion and electron jets in the shock transition region. 
\citet{liu2020} and \citet{wang2020} investigated reconnection in the foreshock. 
\citet{wang2020} found electron-only reconnection in Short Large-Amplitude Magnetic Structures (SLAMS), suggesting that reconnection occurs due to the compression of the SLAMS. 
\citet{liu2020} found electron-only reconnection in shock transients in the ion foreshock. 
Within the transients, $|{\bf B}|$ and $n$ were low and the plasma was turbulent.
{\it MMS} detected high-speed electron jets, enhancements of $\mbox{\boldmath$j$}\cdot\mbox{\boldmath$E'$}$, and enhancements in $T_e$ during the crossings of current sheets in the transient.

\begin{figure}
\begin{centering}
\includegraphics[width=0.8\textwidth]{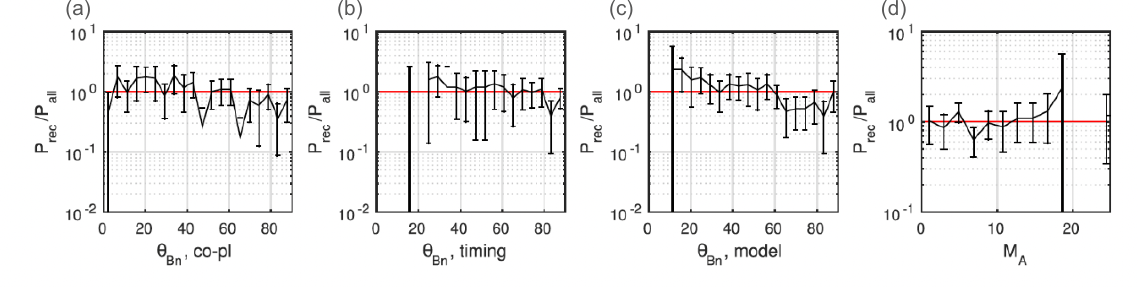}
\caption{Probability to detect reconnecting current sheets in the shock-transition-region/magnetosheath as a function of (a)-(c) shock normal angle $\theta_{Bn}$ using three different methods and (d) Alfv\'en Mach number of the upstream shock \citep[reproduced from][]{gingell2020}.}
\label{fig:SheathRec2}
\end{centering}
\end{figure}

\citet{gingell2020} further statistically studied 223 shock crossings by {\it MMS} - investigating reconnection in the shock transition region of both quasi-parallel and quasi-perpendicular shocks. 
\citet{gingell2020} demonstrated reconnection occurs ubiquitously regardless of shock angle, although quasi-parallel shocks and high Alfv\'en Mach number shocks show slightly higher probability than quasi-perpendicular shocks and low Alfv\'en Mach number shocks as shown in Fig.~\ref{fig:SheathRec2}. 

\subsection{Simulations of Turbulence-Driven Reconnection}\label{sec:TurbulenceDrivenReconnection_sim}

Over the past several decades, numerous numerical simulations of turbulence have demonstrated the process of turbulence-driven reconnection. 
While such simulations are able to reveal vital information about the configuration of turbulence-driven reconnection events that are inaccessible with {\it in situ} observations, the initial identification of reconnection in the simulations presents a challenge, particularly in 3D, which has been a major focus of much of the research. 
Furthermore, the computational challenge of simultaneously resolving all of the collisionless dynamics alongside the large scale separation characteristic of turbulent plasmas, has prompted the exploration of how different plasma models impact turbulence-driven reconnection (see \citet{Shay2024} for a detailed discussion of different numerical plasma models). 

As early as the 1980s, resistive MHD turbulence simulations demonstrated the presence of coherent structures resembling current sheets, where magnetic reconnection can take place~\citep[e.g.][]{Matthaeus1986,Biskamp1989,Politano1989,Carbone1990}.
The characterization and statistical analysis of the distribution and geometry of current sheets
formed in those numerical simulations of MHD turbulence,
as well as their reconnection rates, have been investigated using diverse methodologies~\citep{Dmitruk2004,Servidio2009,Servidio2010,Zhdankin2013,Zhdankin2014,Zhdankin2016a}. 
\citet{Servidio2009,Servidio2010} statistically analyzed reconnection events in 2D MHD turbulence, quantifying, for example, the distribution of reconnection rates and the aspect ratio at each x-point.
In order to find those x-points, they used a topological approach based on classifying the eigenvalues of the Hessian matrix of the vector potential and finding the saddle critical points. A generalization of this method to a 3D geometry for x-points with a guide field, which makes the topological classification more complex, was recently proposed by \citet{Wang2024}.
\citet{Zhdankin2013} developed an algorithm to identify and characterize the geometrical properties of
current sheets in 3D MHD turbulence, and examined how those properties related to the presence of x-points.
\citet{Zhdankin2013} quantified the energy dissipation in 3D MHD turbulence simulations, finding relatively uniformly distributed energy dissipation in the current sheets and also that the number of current sheets increases while their thickness decreases as the magnetic Reynolds number increases.
The relationship between current sheets and other coherent structures was further numerically explored in the 3D MHD simulations of \citet{Zhdankin2016a}.

Similar coherent structures have been identified in simulations using other plasma models, such as two fluid and Hall MHD \citep{Donato2012,Camporeale2018,Papini2019} that can resolve $d_i$ and capture the physics of whistler and kinetic Alfv\'en waves that are thought to play a fundemental role in the sub-proton-scale turbulence in the solar wind and other environments. 
\citet{Donato2012} compared statistics of reconnection events between
Hall-MHD and resistive MHD turbulence simulations. 
Hall-MHD effects led to faster and broader distributions of reconnection rates compared to resistive MHD.
\citet{Camporeale2018} developed a space-filter method to analyze the features of the turbulent cascade due to current sheets in 2D simulations carried out with a two-fluid simulation, which also includes electron inertia. 
Using this approach, \citet{Camporeale2018} was able to quantify how the cascade was modified due to the presence of coherent structures by means of the cross-scale spectral energy-transfer at specific locations in real space.

In contrast to Hall MHD, hybrid-kinetic models also contain ion kinetic physics, which among other effects can describe temperature-anisotropy instabilities and, in general, any process driven by deviations from an equilibrium Maxwell distribution function.
Hybrid-kinetic simulations of turbulence have often been used to investigate and reproduce the ion-scale spectral break and other observed properties of space plasma turbulence~\citep{Franci2015, Franci2020}.
Hybrid-kinetic simulations have revealed the process of current sheet formation and how reconnection can play an important role in regulating
the turbulent cascade near $d_i$~\citep{Cerri2017,Franci2017,Papini2019,Hu2020,Sisti2021,Manzini2023}.
\citet{Papini2019} compared turbulence simulations using Hall-MHD and hybrid-kinetic models, finding good agreement in many turbulent properties between both models. 
\citet{Papini2019} found similar reconnection rates between both plasma models, and by means of quantifying the energy transfer, that there might exist a reconnection-mediated regime at sub-ion scales.
By means of 3D hybrid-kinetic turbulence simulations, \citet{Fadanelli2021} analyzed the energy exchanges near a reconnection site in this scenario. 
They did not determine the energy budget as usually done, but instead a point-to-point analysis that revealed that the conversion to thermal and kinetic energies is statistically related to the local scale of the system, with the largest conversion rate occuring at scales comparable to a few $d_i$.
\citet{Manzini2023} developed a coarse-graining method to measure the nonlinear cross-scale energy transfer at specific locations. 
Applying this method to both {\it MMS} observations in the magnetosheath and hybrid-kinetic turbulence simulations, both cases showed clear indications of a preferential energy transfer to sub-proton scales associated with reconnection. 
\citet{Consolini2023} quantified the fractal dimension of current sheets in 2D hybrid-kinetic turbulence, finding a relationship between this quantity with the spectral features of ${\bf B}$ fluctuations at ion-scales

Other works have developed algorithms to characterize current sheet properties both in 2D and 3D turbulence.
In 2D, \citet{Hu2020} developed such an algorithm based on a convolutional neural network, in which first humans detected reconnection events in turbulence based on several physical signatures including a sufficiently strong ${\bf j}$, ${\bf u}_e$, and ${\bf E}^\prime$ in the out-of-plane direction. 
With this training data, the algorithm automatically identified up to 70\% of the reconnection events in turbulence successfully.
In 3D, \citet{Sisti2021} developed and compared different methods to extract the three characteristic lengths of the formed current sheets.
By comparing with earlier electron-MHD simulations of turbulence by \citet{Meyrand2013}, \citet{Sisti2021} concluded that the additional physics included in the hybrid-kinetic plasma model changed the shape of the most predominant current sheets from ``cigar-like'' to ``knife-like''.

Most hybrid-kinetic simulations have been carried out with a model neglecting electron mass.
As a result, reconnection is driven by physical or numerical resistivity, not by collisionless processes as expected.
In addition, it has been shown that the current sheet width is mainly limited by the grid size~\citep{Azizabadi2021}, so the current sheet thinning process is mainly limited by numerics and not by physical processes.
The affect of electron inertia on current sheets formed by turbulence
was further investigated in 2D hybrid-kinetic simulations by \citet{Jain2022}, which quantified the errors introduced by approximations used in previous hybrid-kinetic codes with electron inertia, concluding that an accurate consideration of electron inertia is important to properly characterize the evolution of electron-scale current sheets in turbulent plasmas.
\citet{Munoz2023} extended this work to 3D, showing how the electron inertia modifies ${\bf j}$, unexpectedly, even at scales $>d_e$ along the direction parallel to ${\bf B}_0$.
In addition, \citet{Munoz2023} emphasized the importance of the electron inertia term in the generalized Ohm's law to balance the reconnection electric field in low-$\beta$ turbulence simulations.
\citet{Califano2020} used another hybrid-kinetic plasma model with electron inertia to investigate the presence of electron-only reconnection in
turbulence. 
The presence of electron-only reconnection was dependent on the wavenumber of the injected fluctuations with more electron-only reconnection events when the fluctuations were injected at wavenumbers comparable to the ion scale, while ion-coupled reconnection appeared when the fluctuations were injected at much larger scales, consistent with the observational work of \citet{Stawarz2022}.
\citet{Arro2020} used a similar code and numerical setup to \citet{Califano2020} in order to determine the influence of electron-only  reconnection on the turbulence.
However, they did not find significant differences between the turbulent fluctuations and intermittency between the cases with electron-only reconnection or ion-coupled reconnection.

Another approach used to investigate turbulence is gyrokinetics.
This is a reduced kinetic model for both electrons and ions that is based on averaging the particle gyromotion and following instead the guiding center of the particles.
The model itself is based on an asymptotic expansion of a small parameter associated with small fluctuations of the distribution function or a small value of the wavenumber anisotropy $k_{\parallel}/k_{\perp}$, among other quantities.
As a consequence, gyroresonances and whistler waves are ruled out of the model, while effects such as electron/ion Landau damping and kinetic Alfv\'en waves are retained.
Turbulence simulations using the gyrokinetic approach have, not only revealed the presence of electron-scale current sheets, but also allowed the quantification of their relative contributions to dissipation~\citep{TenBarge2013,TenBarge2013b,Howes2016}.
For example, it was found that Landau damping plays a fundamental role in dissipating the energy contained in the current sheets.
\citet{Li2023} applied a method to identify current sheets based on a measure of the magnetic flux transport from the separatrices to the reconnection x-line in 3D simulations of gyrokinetic turbulence, revealing that the current sheets formed unexpectedly extended x-lines in the turbulent system.

Fully kinetic turbulence simulations have also shown the presence of both ion and electron-scale current sheets and magnetic reconnection \citep{Karimabadi2013,Karimabadi2014,Wan2012,Wan2015,Wan2016,Haynes2014,Haggerty2017a,Vega2020,Rueda2021,Franci2022,Vega2023}.
\citet{Karimabadi2013} investigated reconnection in 2D fully kinetic simulations of Kelvin-Helmholtz/shear flow generated turbulence, showing how the current sheets are regions with strong localized electron heating due to the parallel reconnection ${\bf E}$.
This heating was found to be much stronger than the heating due to the damping of waves formed in the same system.
\citet{Haynes2014} investigated the relation between reconnection and electron temperature anisotropy via 2D fully kinetic simulations with an implicit PIC code, which allowed them to use a realistic ion-to-electron mass ratio, finding that reconnection sites are associated with strong parallel electron temperature anisotropy, contributing to dissipation.
\citet{Wan2016} compared several 2D and 3D fully kinetic and MHD simulations of turbulence in order to determine the relation between intermittency and dissipation in coherent structures, finding the dissipation measure, ${\bf j}\cdot {\bf E}^\prime - \rho_c{\bf u}_e\cdot{\bf E}$ where $\rho_c$ is the charge density, scales as $\sim |{\bf j}|^2$ in all cases.
\citet{Haggerty2017a} analyzed the statistics of reconnection X-points in 2D fully-kinetic turbulence simulations
by applying similar methods to those of \citet{Servidio2009} for a MHD model.
In contrast to previous MHD simulations, \citet{Haggerty2017a} found that the distribution of reconnection electric fields is broader and can reach up to 0.5 of the local Alfv\'en speed, while keeping an average of 0.1.
\citet{Vega2020} focused on electron-only reconnection in 2D fully kinetic simulation, finding that electron-only reconnection occurs at both high and low electron-$\beta$ with similar reconnection rates. 
\citet{Vega2023} further analyzed 3D fully kinetic simulations of turbulence using an algorithm based on a medial axis transform from image processing , which was capable of identifying and characterizing arbitrary shaped current structures. 
The current structures tended to have half-widths of at most one $d_e$ with a length between $d_e$ and $d_i$. 
Most energy dissipation took place in current structures occupying $\sim20$\% of the total simulation volume and, by identifying large variations in electron flow and characteristic features in the pressure-strain interaction terms, it was estimated that 1\% of current sheets were reconnecting - in contrast to the observational work of \citet{Stawarz2022} which found $\sim10$\% of intense current sheets underwent reconnection, although this discrepancy may owe to different methodologies for identifying distinct current structures. 
Using 3D fully-kinetic anisotropic Alfv\'enic turbulence simulations, \citet{Rueda2021} found that the current sheets generated by the turbulence tended to be less anisotropic than that of the large-scale driving and used several proxies (e.g., $|{\bf j}|$, ${\bf u}_e$, ${\bf u}_i$, and ${\bf E}^\prime$) to identify reconnection sites. 
In a follow-up study, \citet{Rueda2022} further examined the energy transport and dissipation associated with both collionless and effective collision-like terms at the reconnection sites in the simulation. 
\citet{Franci2022} analyzed reconnection events in 2D fully kinetic simulations of turbulence, identifying several reconnection events with a thickness on the order of $d_e$. 
Reminiscent of the work of \citet{Califano2020} and \citet{Stawarz2022}, which looked at the statistical prevalence of electron-only reconnection relative to the dynamics of the driving scale, \citet{Franci2022} found that even within a given simulation turbulent current sheets with shorter lengths tended to appear more electron-only-like, while current sheets with longer lengths appeared ion coupled. 

While many numerical studies of turbulence-driven reconnection have been performed in idealised periodic boxes of turbulence, a number of works, particularly in recent years, spurred on by the {\it MMS} results from Earth's turbulent magnetosheath discussed in Sec.~\ref{sec:TurbulenceDrivenReconnection_obs}, have begun to examine reconnection events generated by turbulence self-consistently excited in shock simulations. 
Reconnection driven by shock turbulence has been studied using kinetic simulations of quasi-perpendicular shocks \citep{Matsumoto2015, Bohdan2020, Lu2021, Guo2023}, quasi-parallel shocks \citep{Gingell2017, Bessho2019, Bessho2020, Bessho2022, Bessho2023, Lu2020, Ng2022, Ng2024}, and across both regimes \citep{Karimabadi2014, Gingell2023, Steinvall2024}.

\subsection{The Role of Reconnection in Turbulent Plasmas}\label{sec:TurbulenceDrivenReconnection_impacts}
Beyond the existence and identification of turbulence-driven reconnection, it is important to consider how and to what extent reconnection contributes to the turbulent dynamics. 
There are several avenues through which turbulence-driven reconnection might impact turbulence, with reconnection potentially i) acting as the dominant nonlinear interaction over some range of scales and ii) contributing to the dissipation of the turbulence. 

\subsubsection{Impact on the Energy Spectrum. }
As discussed in Sec.~\ref{sec:theory_spectra}, theoretical descriptions of turbulence are built on assumptions about the physical interactions controlling the nonlinear dynamics. 
Several works have explored how to incorporate reconnection into this framework, with early work basing the analysis on the isotropic weak turbulence formalism of Iroshnikov-Kraichnan \citep{Carbone1990}, while recent works have examined the topic using anisotropic MHD \citep{Loureiro2017a, Mallet2017a} and collisionless \citep{Loureiro2020, Mallet2020} turbulence. 

In the theoretical scenarios proposed in these works, turbulent dynamics form current-sheet-like structures with aspect ratios (length relative to thickness) that become progressively more anisotropic at smaller scales in a manner consistent with the spectral anisotropy predicted by an anisotropic turbulence model.  
The tearing instability is assumed to initiate reconnection at these current sheets once the aspect ratio becomes sufficiently large and alters the turbulence by ``disrupting''/``destroying'' the elongated current sheets over the timescale of the tearing instability growth, thereby altering the distribution of energy in spectral space. 
While the tearing instability growth is treated as a linear instability, it is both initiated via the formation of a nonlinear structure and leads to the development of a fully nonlinear perturbation to the current sheet and, in this sense, the linear growth rate is taken to characterise the rate at which the nonlinear reconnection dynamics develop. 
If a range of scales exists over which this tearing timescale is faster than the dynamical timescales of other nonlinear effects, such as those generating the current sheets, then it is supposed that the tearing timescale will be the relevant timescale to associate with $\tau_{tr}$. 
This picture relies on reconnection being sufficiently prevalent so as to make a significant impact, the scale-dependant anisotropy of current structures reflecting the spectral anisotropy of the turbulence model, and the linear tearing instability being the correct way to characterize the initiation of reconnection in a turbulent environment.  

Based on this scenario, predictions for two key parameters can be derived - the critical scale ($a_c$) at which reconnection becomes the dominant nonlinear interaction and the power law scaling $\mathcal{E}\left({\bf k}\right)$.
Constraining these parameters requires both a model for the large-scale turbulent dynamics setting up the current sheets and a description of the tearing instability, which depends on whether the system is resistive or collisionless and on the exact profile of the current sheets. 

For resistive MHD, several 3D anisotropic turbulence models have been proposed, including the dynamic alignment model of \citet{Boldyrev2006} and intermittency models of \citet{Chandran2015} and \citet{Mallet2017b}. 
\citet{Loureiro2017a} and \citet{Mallet2017a} demonstrated that for these models, at sufficiently large $Re$, a range of scales exists where the resistive tearing instability is faster than the nonlinear dynamics generating the current sheets, suggesting the presence of a reconnection mediated inertial range.
For the case of a hyperbolic tangent current sheet profile, $a_c$ was found to be
\begin{equation}
a_c/\lambda_C\sim\left(V_{A,\lambda_C}\lambda_C/\eta\right)^{-4/7},
\end{equation}
where we have identified $\lambda_C$ with the outer scale of the turbulence and $V_{A,\lambda_C}$ is the Alfv\'en speed for fluctuations at the outer scale; and $\mathcal{E}\left(k_\perp\right)\sim k_\perp^{-11/5}$ in the reconnection-mediated range. 
Recent simulations of high-$Re$ MHD turbulence in 2D \citep{Dong2018} and 3D \citep{Dong2022} have provided evidence for the presence of reconnection and the expected steepening of the inertial range spectrum at $a_c$. 

Collisionless effects both alter the tearing instability, as well as the nonlinear dynamics of the turbulence. 
Two scenarios, one where $a_c$ is larger than the ion scales and one where $a_c$ is smaller than the ion scales, can potentially occur. 
In the former scenario, the analysis proceeds in a similar manner to the resistive case, but with a modified expression for the tearing instability leading to 
\begin{equation}
a_c/\lambda_C\sim\left(d_e/\lambda_C\right)^{4/9}\left(\rho_s/\lambda_C\right)^{4/9}
\end{equation}
where $\rho_s$ is the ion acoustic scale, and $\mathcal{E}\left(k_\perp\right)\sim k_\perp^{-3}$ is obtained for a hyperbolic tangent current sheet profile \citep{Loureiro2017b, Loureiro2020}.
The range of spectral indices found for different current sheet profiles is reminiscent of those reported in the so-called transition range of solar wind turbulence, although reconnection is not the only explanation that can produce such slopes \citep{Bowen2020}. 
The later scenario, where $a_c$ occurs in the kinetic scales, is more challenging due to less well understood anisotropic turbulence models; however, some work has been done on this scenario \citep{Loureiro2017b, Boldyrev2019, Mallet2020, Loureiro2020, Boldyrev2020}, which may be relevant for understanding the impact of electron-only reconnection. 

\subsubsection{Contribution to Energy Dissipation. }
Magnetic reconnection can also facilitate the dissipation of turbulence. 
Simulations \citep{Servidio2011} and observations \citep{Osman2012, Chasapis2017b, Chasapis2018b} suggest intermittent structures are locations of enhanced temperature and energy conversion. 
As discussed in Sec.~\ref{sec:TurbulenceDrivenReconnection_obs}, missions such as {\it Cluster} and {\it MMS} have enabled the identification of thin reconnection events and the direct examination of the local energy conversion associated with them, which may account for nontrivial amounts of energy conversion when compared with estimates of the overall energy budget \citep{Sundkvist2007, Schwartz2021}. 
However, other studies, such as that by \citet{Hou2021}, which examined integrated ${\bf j}\cdot{\bf E}^\prime$ at intense PVI structures, while simultaneously identifying PVI structures associated with reconnection, concluded that, while reconnection events may have large energy conversion signatures, their integrated contribution to energy dissipation may be small ($\sim15$\% of the dissipation associated with large PVI structures and $\sim1$\% of the total integrated ${\bf j}\cdot{\bf E}^\prime$ in the analysed interval) due to the small size of the diffusion region and limited occurrence rate.

One limitation of local analyses of energy conversion, is that, while the diffusion regions, which contain some of the strongest gradients, occupy a small volume; the entire volume of the reconnection outflows and separatrices, occupy a much larger volume and can also be energetically important for the particle acceleration and heating. 
Furthermore, since reconnection involves the inflow of particles from the surrounding environment, in can lead to the acceleration and heating of a larger effective volume of particles than expected from the size of the current sheet alone.

An alternative way to assess the importance of reconnection for turbulent dissipation is to consider the energy budget of the reconnection events generated by the turbulent dynamics. 
The amount of magnetic energy released by reconnection that is available to each electron-proton pair is given by
\begin{equation} \label{eq:rec_energy}
\mathcal{E}_{rec} = m_i V_{A,inflow}^2, 
\end{equation} 
where $m_i$ is the ion mass and $V_{A,inflow}$ is the Alfv\'en speed associated with the inflowing reconnecting component, $B_L$, of the magnetic field. 
Taking into account the potential effect of asymmetry on either side of a reconnecting current sheet, $V_{A,inflow}$ is given by
\begin{equation} \label{eq:VA_inflow}
V_{A,inflow} = \sqrt{ \frac{\left|B_{L,1}\right|  \left|B_{L,2}\right| \left( \left|B_{L,1}\right| + \left|B_{L,2}\right| \right)}{\mu_0 m_i \left(n_1 \left|B_{L,2}\right|  + n_2 \left|B_{L,1}\right|  \right)}},
\end{equation} 
with subscripts $1$ and $2$ denoting values on either side of the current sheet. 
The total rate of energy dissipation associated with reconnection can be quantified by taking into account the fraction of $\mathcal{E}_{rec}$ converted into particle heating (or energetic particle acceleration) and the rate that magnetic flux is reconnected. 
Denoting the net rate of energy dissipation per unit mass associated with reconnection within a turbulent volume as $\epsilon_{rec}$, which will be equivalent to $\epsilon$ if reconnection accounts for all of the dissipation in the volume, gives \citep{Shay2018, Stawarz2022}
\begin{equation}\label{eq:rec_dissipation_rate}
    \epsilon_{rec} = \sum_j f_{rec,j} \alpha_j V_{A,inflow,j}^2 \left(\frac{V_{A,inflow,j}}{\lambda_j} \mathcal{R}_j \right). 
\end{equation}
In Eq.~\ref{eq:rec_dissipation_rate}, $f_{rec}$ is the fraction of particles in the turbulent volume processed by a given reconnection event, $\alpha$ is the fraction of $\mathcal{E}_{rec}$ converted into particle heating, $\lambda$ is the length of the inflow region, and $\mathcal{R}$ is the dimensionless reconnection rate, such that $\mathcal{R}V_{A,inflow}/\lambda$ represents the inverse timescale over which magnetic flux is reconnected. 
The summation over $j$ represents a sum over each reconnection event in the turbulent region at any given time, such that subscript $j$ denotes a quantity for a given reconnection event.

While Eq.~\ref{eq:rec_dissipation_rate} can be straightforwardly computed if all reconnection events within a volume can be characterized, in many cases this is not possible and it can be beneficial to estimate Eq.~\ref{eq:rec_dissipation_rate} based on characteristic values for the reconnection events, such that
\begin{equation}\label{eq:rec_dissipation_rate2}
    \epsilon_{rec} \sim N_{rec} f_{rec} \alpha V_{A,inflow}^2 \left(\frac{V_{A,inflow}}{\lambda} \mathcal{R} \right), 
\end{equation}
where $N_{rec}$ is the number of reconnection events in a turbulent volume. 
Rough estimates of the above parameters constrained by the reconnection events observed in the systematic survey of reconnection in the turbulent magnetosheath by \citet{Stawarz2022} obtained dissipation rates of $1\times10^4$ to $3\times10^6$ J/kg-s, assuming $\lambda\sim\lambda_{C,{\bf B}}$ consistent with reconnection occuring at the interface of correlation length magnetic structures. 
While there was a large spread, these estimates were in rough agreement with previous independent estimates of $\epsilon$ in the magnetosheath obtained from expressions similar to Eq.~\ref{eq:yaglom} \citep{Hadid2018, Bandyopadhyay2018}, suggesting reconnection is a non-trivial contributor to dissipation - potentially alongside other processes. 

\citet{Shay2018} derived a related expression for the energy dissipation associated with reconnection, which can be thought of as an extension to Eq.~\ref{eq:rec_dissipation_rate2} that parameterize $\alpha$ based on expectations from guide field reconnection. 
Based on a series of 2D laminar symmetric guide field reconnection simulations \citet{Shay2018} find that ion and electron heating is well parameterized by 
\begin{eqnarray}
\Delta T_i &=& c_i\left(\frac{|B_L|}{|{\bf B}|}\right)^2 m_i V_{A,inflow}^2 \label{eq:shay_heatingi} \\
\Delta T_e &=& c_e\left(\frac{|B_L|}{|{\bf B}|}\right) m_i V_{A,inflow}^2 \label{eq:shay_heatinge}
\end{eqnarray}
where $\Delta T_{i,e}$ are increases in ion and electron temperatures and $c_{i,e}$ are constants of proportionality associated with ion and electron heating, respectively. 
While these expressions were empirically derived, Eq.~\ref{eq:shay_heatingi} is consistent with theoretical expectations for ion acceleration in contracting magnetic islands \citep{Drake2009}. 
Based on Eqs.~\ref{eq:shay_heatingi} and \ref{eq:shay_heatinge}, $\alpha_j = c_i\left(|B_{L,j}|/|{\bf B}_j|\right)^2 + c_e\left(|B_{L,j}|/|{\bf B}_j|\right)$ in Eq.~\ref{eq:rec_dissipation_rate}. 
Using this value of $\alpha_j$, Eq.~\ref{eq:rec_dissipation_rate} is then consistent with Eq.~3 of \citet{Shay2018} divided by the total mass in the turbulent volume, noting that $f_{rec}$ is equivalent to the volume of magnetic island that have reconnected in the \citet{Shay2018} formalism divided by the total volume of the region. 
Due to the different scaling of the ion and electron heating, the relative ion to electron heating varies depending on the distribution of local guide field strength at turbulence-driven reconnection sites. 
Relating the expressions for the ion and electron heating rates to the properties of the turbulent fluctuations in a heuristic manner and comparing with the heating rates obtained from 2.5D PIC simulations of turbulence, \citet{Shay2018} found that, while the individual scalings for ion and electron heating did not agree with the prediction, the ratio did agree well.

\section{Reconnection-Driven Turbulence} \label{sec:ReconnectionDrivenTurbulence} 

The converse of the scenario discussed in Sec.~\ref{sec:TurbulenceDrivenReconnection}, referred to here as reconnection-driven turbulence, in which large-scale reconnection sites set up by system-scale dynamics act to generate turbulent dynamics, has also been an active area of investigation. 
As discussed in \citet{Graham2024}, reconnection is known to generate a variety of waves in the diffusion regions, along the separatrix, and in the exhausts that are excited by the free energy available in non-Maxwellian distributions or strong gradients. 
As these waves grow to large amplitudes and nonlinearly interact, they can lead to turbulence. 
The plasmoid/tearing instability can also destabilize the reconnecting current sheet and spontaneously generate multiple x-lines with plasmoids (or flux ropes in 3D) between them. 
These plasmoids/flux ropes can interact and merge in the outflows, producing a complex turbulent character to the current layer \citep{Daughton2011, Oishi2015, Huang2016t}. 
Turbulent dynamics also can be generated by the interaction of the exhaust with its surroundings, such as through shear instabilities or the exhaust encountering an obstacle. 

Given the relatively large amount of data, spanning many correlation lengths, needed to perform typical turbulence analyses, relatively limited observational analyses of the fluctuations within reconnection exhausts have been performed in the solar wind \citep{Miranda2021, Eastwood2021, Wang2023} and at Earth's magnetopause \citep{Ergun2017}, which suggest enhanced turbulent fluctuations within the exhausts. 
However, Earth's magnetotail has provided an ideal environment for the observational examination of reconnection-driven turbulence. 
This status is partially associated with the central role that system-scale Dungey-Cycle-like reconnection plays in energy transport in the magnetotail and the need to understand the role of turbulence in mediating this transport. 
Additionally, due to the nature of the system, the effective spacecraft trajectory through a reconnection event in the magnetotail is often such that significant dwell times in the reconnection exhaust are obtained, which is conducive to turbulence analyses (in contrast to solar wind or magnetopause reconnection encounters where significant and sustained components of the background motion normal to the current sheet, combined with relatively narrow outflow thicknesses, mean that rapid transits of the outflow are the norm). 
Although it may be possible to identify additional regions in near-Earth space, such as the heliospheric current sheet in the inner heliosphere, where extended dwell times are also possible. 

\subsection{Magnetotail Turbulence} \label{sec:ReconnectionDrivenTurbulence_obs}
As illustrated in Fig.~\ref{fig:turbulent_magnetosphere}, the stretched magnetic field in Earth's magnetotail can be broadly divided into two regions - the northern and southern lobes featuring low $n$ and strong $|{\bf B}|$; and the relatively high density plasma sheet surrounding the ${\bf B}$ reversal at the center of the magnetotail. 
The plasma sheet is a dynamic region and early studies examined the fluctuations in the plasma sheet as a whole, demonstrating the region exhibits nonlinear behavior consistent with turbulence albeit with differences relative to classical homogeneous turbulence potentially associated with boundary effects, coupling to Earth's ionosphere, and nonuniform driving \citep{Borovsky1997, Borovsky2003, Weygand2005}.

In the plasma sheet, system-scale reconnection occurs at x-lines near Earth at $\sim25R_E$ and in the distant tail at $>60R_E$ from Earth \citep[see][]{Fuselier2024}, driving tailward and Earthward exhausts.  
These outflows can drive turbulent fluctuations in the plasma sheet and subsequent analyses demonstrated such flows contain some of the clearest and most intense signatures of turbulence in the plasma sheet \citep{Bauer1995, Voros2004, Voros2006, Stawarz2015}. 
Reconnection outflows in the magnetotail are typically transient and likely linked with Bursty Bulk Flows (BBFs) in observations. 
As the flows plow through the surrounding plasma, they relax the stretched $B_{x,GSM}$ component of the field in Geocentric Solar Magnetic (GSM) coordinates and enhance the dipolar $B_{z,GMS}$ component producing so-called Dipolarization Fronts. 
On the Earthward side, as the flows impinge on the strong nearly dipolar near-Earth field, the flows are slowed and deflected generating large-scale vorticies \citep{Panov2010}. 

\begin{figure}
\begin{centering}
\includegraphics[width=0.6\textwidth]{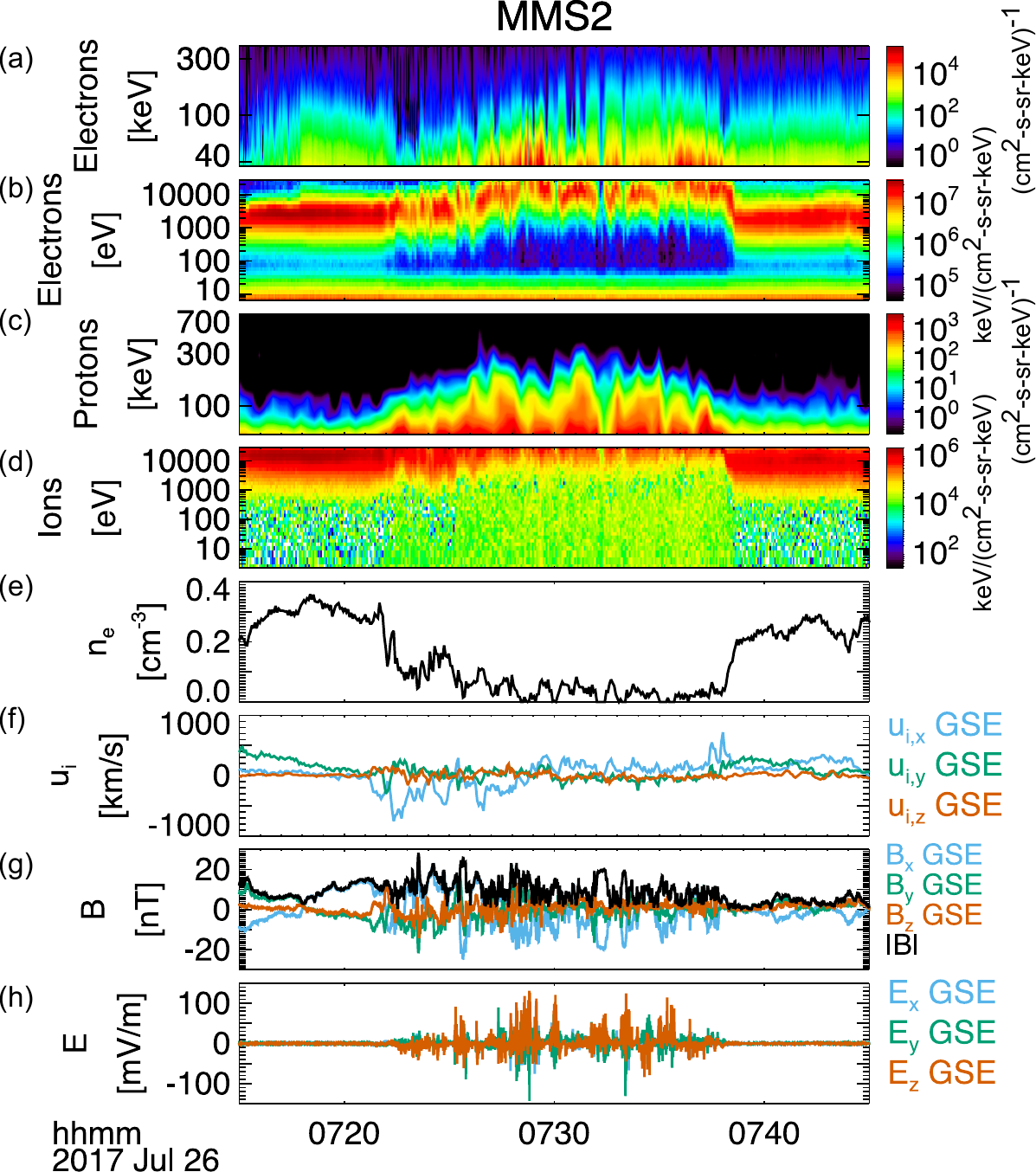}
\caption{Example turbulent x-line crossing observed by {\it MMS}, showing (a) high-energy omnidirectional electron energy fluxes, (b) low-energy omnidirectional electron differential energy fluxes, (c) high-energy omnidirectional proton energy fluxes, (d) low-energy omnidirectional ion differential energy fluxes, (e) electron number density, (f) ion flow velocity, (g) ${\bf B}$, and (h) ${\bf E}$. All vector quantities are in Geocentric-Solar-Ecliptic (GSE) coordinates.}
\label{fig:turbulent_magnetotail}
\end{centering}
\end{figure}

{\it Cluster} and {\it MMS} have provided clear examples of extended encounters with reconnection x-lines and adjacent outflows featuring exceptionally intense turbulent fluctuations \citep{Ergun2018}, as well as tailward and Earthward exhausts likely further from the x-line with clear evidence of turbulent dynamics \citep{Voros2004, Eastwood2009}. 
Fig.~\ref{fig:turbulent_magnetotail} illustrates a notable example from 26 July 2017 where {\it MMS} continually observed a near-Earth x-line featuring exceptionally intense fluctuations for 20 minutes. 
While the event has intense fluctuations in all quantities, an overall variation from $B_{z,GSE}>0$ to $B_{z,GSE}<0$ in conjunction with a variation from $u_{x,GSE}<0$ to $u_{x,GSE}>0$ during the encounter, consistent with {\it MMS} traversing through the x-line from the tailward to Earthward outflows. 
An overall variation from $B_{x,GSE}>0$ to $B_{x,GSE}<0$ also occurs, consistent with {\it MMS} concurrently traveling from southward to northward of the current sheet. 
In this and other similar events, $n$ is exceptionally low, suggesting field lines carrying dense plasma from the plasma sheet have already reconnected and been evacuated from the region, and now lobe field lines are reconnecting. 
\citet{Ergun2020a} suggested the large amount of energy imparted to each particle in the reconnection region ($\sim3$keV/s per proton/electron pair in this event), owing to the low $n$ and large incoming Poynting flux from the lobes, may be a key reason why such strongly turbulent fluctuations are generated in this type of event.  

${\bf B}$ and ${\bf E}$ spectra in these turbulent x-lines and outflows show broadband power law scalings across MHD, sub-proton, and electron scales consistent with theory and observations of other turbulent environments \citep{Voros2004, Eastwood2009, Ergun2018, Ergun2020a}. 
At sub-proton scales, the fluctuations appear consistent with kinetic Alfv\'en waves, the dissipation of which may lead to significant proton heating \citep{Chaston2012, Chaston2014}. 
Due to the variable flows in the magnetotail, even within fast reconnection outflows, some care needs to be taken in the analysis of spectra and other scaling properties, since the Taylor hypothesis is likely not strictly valid. 
Based on a careful analysis of multipoint timing velocities of ${\bf E}$ and ${\bf B}$ structures in the turbulent x-line shown in Fig.~\ref{fig:turbulent_magnetotail}g--h, \citet{Ergun2020a} found that, while the direction of propagation was random, the speed of structures were typically between the Alfv\'en and ion acoustic speeds and within a factor of two of each other. 
Using the average speed as a conversion between frequency and wavenumber, \citet{Ergun2020a} found good agreement between the location of spectral break points and characteristic plasma length scales, as well as good correspondence between spatial and temporal correlation functions. 
\citet{Bandyopadhyay2021} examined Eq.~\ref{eq:yaglom} using {\it MMS} measurements of a reconnection outflow on 16 June 2017, providing evidence of an active cascade of energy to small scales within the reconnection exhaust.
\citet{Jin2022} examined higher-order turbulence statistics in another turbulent x-line encountered by {\it MMS} on 28 May 2017, finding evidence of turbulent intermittency. 
By dividing the extended reconnection outflow into subintervals, \citet{Jin2022} found evidence of increasingly intermittent behaviour with distance from the x-line, suggesting an evolution of the turbulent dynamics through the outflow region. 
Recently, \citet{Richard2024} performed a statistical survey of turbulence in BBFs observed from 15--25$R_E$ downtail of Earth by {\it MMS}, which confirmed many of the results inferred from case studies of magnetotail reconnection jets. 
Importantly, it was determined that turbulence quickly develops within a few ion gyroperiods with a driving scale comparable to the size of the jet. 
An intense energy cascade rate (relative to the solar wind or magnetosheath) was also identified, which extend over an order of magnitude in scale. 

While somewhat rare, in several turbulent x-line events, EDRs have been encountered and identified using {\it MMS} data. 
Despite the strong fluctuations, the basic properties and structure of the EDR appears broadly consistent with quasi-2D laminar reconnection and the reconnection process continues amidst the turbulent fluctuations for an extended duration \citep{Ergun2022, Qi2024}. 
\citet{Ergun2022} found by analysing one event observed on 27 August 2018 that the typical features of reconnection (e.g., a persistent ion jet, ${\bf B}$ profile including a $B_L$ reversal and Hall fields) were all present as expected, but with additional fluctuations on top of them. 
Off-diagonal terms of the electron stress tensor, which encode off-diagonal contributions to generalised Ohm's law from both the electron pressure and electron inertial effects, were found to both account for the observed reconnection electric field and could be understood from a laminar reconnection perspective. 
\citet{Qi2024} examined another event, noting that, while large reconnection electric fields may be present, the overall aspect ratio of the event, which provides another estimate of the reconnection rate, gives a value of $\sim0.2$ consistent with typical estimates of ``fast'' reconnection rates. 

While turbulence may not significantly alter the electron dynamics at the x-line itself, at least for the few examples that have lent themselves to detailed examination, they may contribute to repartitioning energy in the exhausts.
As seen in Fig.~\ref{fig:turbulent_magnetotail}a and c, intermittent bursts of energetic ions and electrons are interspersed with the fluctuations. 
\citet{Ergun2018} found that both structures with ${\bf j}\cdot{\bf E}^\prime$ dominated by the components parallel and perpendicular to ${\bf B}$ contribute to the turbulent dissipation. 
${\bf j}_{\perp}\cdot{\bf E}_\perp^\prime$ contributed to 80\% of the net dissipation and primarily acted at frequencies near the ion cyclotron frequency, while ${\bf j}_{||}\cdot{\bf E}_{||}^\prime$ primarily acted at higher frequencies and led to the acceleration of some of the most energetic electrons. 
Further analyses using observations and test particle models demonstrated that magnetic deletions associated with the intense fluctuations enabled particle trapping in regions of energy conversion \citep{Ergun2020b, Ergun2020a}. 
Notably, perpendicular energization tended to occur making it more difficult for particles to escape, thus leading to further energization and the generation of the non-thermal energetic particles seen in the observations. 
\citet{Bergstedt2020} performed a statistical analysis of magnetic structures in the same event finding ${\bf j}_{||}\cdot{\bf E}_{||}^\prime$ dissipation occured at current sheets at the interface of plasmoid-like structures, which may be consistent with the dynamic production and merger of plasmoids by the reconnecting current sheet. 
\citet{Oka2022} examined particle energization, both in reconnection events like the example in Fig.\ref{fig:turbulent_magnetotail} where lobe field lines reconnect and in events where plasma sheet field lines carrying denser plasma reconnect, demonstrating larger ${\bf E}$ fluctuations are present in lobe reconnection. 
Interestingly, the stronger ${\bf E}$ fluctuations were linked to a smaller nonthermal energy fraction. 
This discrepancy with \citet{Ergun2020b} may be linked to differences in the definition of nonthermal particles, with \citet{Oka2022} focusing on the high-energy power law tail to the particle distributions, while \citet{Ergun2020b} focused on a high-energy non-thermal shoulder to the distribution. 

Turbulent fluctuations can also be excited through the interaction of the exhaust with the surroundings.  
\citet{Volwerk2007} found evidence of the Kelvin-Helmholtz instability associated with the velocity shear along the edges of a BBF using a conjunction between {\it Cluster} and {\it DoubleStar} and \citet{Divin2015} found evidence of turbulence caused by the LHDI at the dipolarization front.
Closer to Earth in the BBF braking region, studies have identified broadband turbulent spectra of kinetic-Alfv\'en-wave-like fluctuations, the excitation of which is likely partially associated with the flow braking process and may play a role in the conversion of bulk flow kinetic energy into other forms \citep{Chaston2012, Ergun2015, Stawarz2015}. 
\citet{Ergun2015} and \citet{Stawarz2015} found using {\it THEMIS} data that the BBF braking region was filled with large-amplitude high-frequency $E_{||}$ structures such as double layers and electron phase space holes, potentially excited by field-aligned current instabilities associated with the turbulence. 
\citet{Stawarz2015} estimated these $E_{||}$ structures could play a significant role in the dissipation of the turbulent fluctuations and heating of the plasma; and subsequent large-scale statistical surveys have supported the picture of the braking region as a region of enhanced solitary wave activity, as well as proton and electron heating \citep{Hansel2021, Usanova2022, Usanova2023}. 
Based on a statistical examination of BBF events, \citet{Chaston2014} found an enhanced divergence of field-aligned Poynting flux associated with kinetic Alfv\'en waves away from the center of the plasma sheet for distances $<15R_E$ downtail of Earth, suggesting a fraction of the small-scale fluctuations excited in the region propagate along the field-lines depositing energy in the auroral region. 
In this scenario, the turbulent dynamics, while not responsible for dissipating the energy radiated from the region, convert bulk flow energy into small-scale fluctuations capable of propagating to the ionosphere.  
\citet{Stawarz2015} estimated the energy budget in the braking region, finding turbulent energy dissipation, adiabatic heating due to the compression of the magnetic field in the braking region, and radiated Poynting flux were similar in magnitude and the sum of these energy losses were comparable to the energy input into the region due to the bulk flows. 
At the inner edge of the flow braking region around $\sim7R_E$ downtail of Earth, BBFs have been associated with enhanced turbulence and the dissipation of these fluctuations has been implicated in seeding the outer radiation belt \citep{Ergun2022b}. 

\subsection{Simulations of Reconnection-Driven Turbulence} \label{sec:ReconnectionDrivenTurbulence_sims}

Numerical simulations have also long been used to examine the turbulence that is self-generated by magnetic reconnection under certain circumstances. 
Arguably, most of the numerical work on reconnection-driven turbulence has been performed using MHD plasma models, for example in the context of the plasmoid instability or other secondary fluid instabilities; however, the impact of kinetic effects and instabilities more appropriate to the collisionless space plasmas of the near Earth environment have been historically less well understood. 
In particular, turbulence near and in the diffusion regions has been a critical area of research due to its potential role in balancing the reconnection electric field, and thereby enabling reconnection, through anomalous resistive and viscous effects that can potentially enhance the reconnection rate. 
Such anomalous effects arise from the nonlinear contributions to generalized Ohm's law due fluctuating quantities that are not typically treated in the traditional quasi-laminar reconnection picture.
Note that this way in which turbulence affects reconnection is somewhat different from the stochastic reconnection picture, often analysed in the MHD context and discussed further in Sec.~\ref{sec:StocasticReconnection}.
In this section, we discuss a selection of the most important findings from the past several years, with a focus on instabilities causing turbulence in kinetic reconnection simulations applied to space plasmas. 

In 2D, several works have shown streaming instabilities between different electron and/or ion populations can generate turbulence, particularly at the separatrices that are known to host a variety of waves due to instabilities and other nonlinear mechanisms \citep{Fujimoto2014,Goldman2014}. 
\citet{Cattell2005} and \citet{Pritchett2005} performed 2D fully kinetic PIC simulations of magnetic reconnection with a guide field and parameters appropriate to Earth's magnetotail. 
This configuration produced electron holes and associated turbulence via streaming (Buneman) instabilities at the separatrices, consistent with in-situ observations.
Later simulations with more realistic parameters, such as a realistic ion-to-electron mass ratio and larger domain, demonstrated this process occurs for all guide fields and is, therefore, expected under a variety of conditions in Earth's magnetosphere \citep{Lapenta2011a}.
\citet{Munoz2016} showed using 2D fully kinetic PIC simulations that there is a strong guide field regime in which broadband electrostatic turbulence develops in the separatrices and outflows, with the key finding that the separatrix/outflow turbulence is associated with double-peaked and anisotropic electron distribution functions and wave activity near the lower-hybrid frequency whose strength is correlated with the instantaneous reconnection rate.
This turbulence was shown to be associated with anomalous resistivity, which was smaller than the electron inertia or non-gyrotropic electron pressure tensor in generalised Ohm's law \citep{Munoz2017}. 
However, since the process was not occurring at the x-line, it was not associated with enabling the reconnection process itself. 
Using 2D fully-kinetic simulations with a sufficient scale separation between $d_e$ and the Debye length (a parameter that is typically small in PIC simulations of reconnection), \citet{Jara-Almonte2014} showed that Debye-scale turbulence was excited near the x-line as well, through the action of streaming instabilities in the reconnection plane. 

3D fully kinetic reconnection simulations tend to be more turbulent than their 2D counterparts due to the presence of the additional degree of freedom enabling more instabilities with wavenumbers along the additional dimension. 
\citet{Drake2003}, \citet{Che2011}, and \citet{Che2017a} found that in 3D the current layer develops filamentary magnetic structures due to electron-shear flow and Buneman instabilities along the direction parallel to the current sheet using PIC simulations with a relatively strong guide field, small spatial domain, and initially force-free equilibrium. 
During the non-linear phase of these instabilities, electron holes and turbulence were generated, which produced anomalous resistivity and viscosity.
\citet{Daughton2011} carried out 3D PIC simulations but with a smaller guide field, much larger spatial
domain, and initialized with a Harris current sheet equilibrium.
In these simulations, the current sheet developed flux ropes at electron scales due to the tearing instability, which were not seen in previous simulations because of their smaller spatial domain.  
These flux ropes underwent  complex 3D interactions that lead to continuously self-generated and inhomogeneous turbulence within the electron current layer.
Later simulations by \citet{Liu2013} revealed that the turbulence generated under similar conditions to \citet{Daughton2011} does not modify the reconnection rate via anomalous resistivity or viscosity, in contrast to the earlier results by \citet{Drake2003} and \citet{Che2011}.
The discrepancy was explained as being due to the lack of streaming or electron shear flow instabilities in the higher plasma-$\beta$ regime appropriate to magnetospheric conditions and enhanced parallel heating that can be developed in simulations with larger spatial domains.
\citet{Fujimoto2021, Fujimoto2023} demonstrated that in 3D simulations without a guide field a different scenario occurs, in which electromagnetic turbulence is generated by an electron Kelvin-Helmholtz instability that produce. 
These fluctuations are capable of producing anomalous viscosity due to electron transport that is capable of breaking the electron frozen-in condition, but not anomalous resistivity since the turbulence mainly effects the electrons, while the ions remain decoupled. 

Another source of free energy to drive turbulence is the lower-hybrid drift instability (LHDI) generated by gradients in the density or magnetic field associated with diamagnetic currents. 
This instability can be particularly important in the case of asymmetric reconnection, such as at Earth's dayside magnetopause, where the strong density gradients across the reconnecting current sheet are conducive to exciting the instability. 
The LHDI typically generates waves initially at the edge of the current sheet that can spread toward its center.
Turbulence associated with the LHDI has been historically considered as a source of anomalous resistivity. 
However, only in the last decade have 3D fully-kinetic PIC simulations with a large-enough scale separation between electron and ion scales been able to reveal under which conditions this instability can modify the reconnection properties.
\citet{Roytershteyn2012} and \citet{Pritchett2012} found that the LHDI can cause enough turbulence near the current sheet center to sustain the reconnection electric field if the plasma-beta is low enough under asymmetric conditions, although most of the LHDI-driven turbulence is confined to the separatrices. 
Such conditions are unlikely to be met for magnetopause reconnection.
Later similar simulations obtained different conclusions under different conditions.
\citet{Price2016, Price2017d} carried out 3D fully kinetic simulations of an asymmetric magnetopause reconnection event based on observations made by {\it MMS}. 
Unlike \citet{Roytershteyn2012}, these simulations found LHDI-driven turbulence at both the x-line and sparatricies that was strong enough to balance the reconnection electric field. 
The discrepancy was attributed to different boundary conditions and a stronger than previously expected density jump across the current sheet in the observed {\it MMS} event. 
Although, \citet{Le2017, Le2018} modelled the same reconnection event as \citet{Price2016, Price2017d}, concluding that anomalous resistivity was small, but that the turbulence acted to enhance plasma mixing and heating in the event.
The simulations by \citet{Price2016} also revealed that crescent-shaped electron distribution functions, an important hallmark of magnetopause reconnection often observed by {\it MMS}, were not affected by the turbulence developed in the current sheet. 
Later, \citet{Price2020} explored a similar system but under the presence of a significant guide field, showing that in this case turbulence develops in the diffusion region due to a variant of the LHDI.  
While the anomalous resistivity produced by the electromagnetic fluctuations at the x-line were small, other anomalous terms were significant, but did not significantly impact the reconnection rate. 
{\it MMS} observations of lower hybrid fluctuations associated with reconnection at the magnetopause are often consistent with the properties found in 3D simulations \citep[for a detailed overview of the LHDI associated with reconnection from spacecraft observations and simulations see][]{Graham2024} and feature broadband power-law spectra suggestive of turbulent dynamics. 
However, such fluctuations (both in observations and simulations) are often observed in narrow boundary regions, making it challenging to apply typical turbulence analyses and it remains unclear whether nonlinear processes play a significant or dominant role in the evolution of the waves or if there is significant energy transfer across spatial scales. 

As discussed in Sec.~\ref{sec:ReconnectionDrivenTurbulence_obs}, turbulence is also generated in the reconnection outflows. 
\citet{Pritchett2010a} and \citet{Vapirev2013} showed that in 3D fully kinetic simulations a Rayleigh-Taylor-like interchange instability occurs in association with the density gradients at the dipolarization front, generating a turbulent outflow. 
\citet{Lapenta2015} investigated the outflows using similar simulations, finding signatures of secondary reconnection events embedded within the complex turbulent fluctuations in the outflows. 
Subsequent works were able to automatically identify such secondary reconnection events using machine learning methods \citep{Lapenta2022}. 
Numerical simulations have also demonstrated the ability of turbulence in the outflows to efficiently accelerate electrons into non-thermal power-law tails similar to those observed in the magnetotail observations \citep{Lapenta2020}.

For extended (long) current sheets, the plasmoid instability is also well-known to occur in numerical simulations. 
In the context of MHD, theory predicts a long current sheet can generate a chain of secondary magnetic islands with secondary current sheets between them if the Lundqvist number ($S=\mathcal{L}V_A/\eta$; i.e., similar to the magnetic Reynolds number but based on the Alfv\'en velocity) is large enough relative to the aspect ratio of the current sheet \citep{Shibata2001,Loureiro2007}. 
In this regime, the reconnection rate is expected to be independent on resistivity.
Each secondary current sheet can also be plasmoid-unstable with possibly leading to a downward cascade of plasmoids in a fractal way until they eventually reach kinetic scales, triggering kinetic reconnection.
Turbulence arises due to the interaction between those plasmoids (or flux ropes in 3D).
This plasmoid instability has received significant attention and has been mainly analyzed using MHD simulations \citep[see, e.g.,][and references therein]{Barta2011a,Huang2016t}.
The transition from a collision-dominated plasmoid instability to kinetic reconnection was studied in 3D PIC simulations including a collision operator \citep{Stanier2019}.
But, in general, there has been relatively little work of collisionless kinetic reconection simulations of plasmoids with a focus on the self-generated turbulence.
One relevant example of such a work is \citet{Fujimoto2012}, who simulated 3D reconnection with a long current sheet, finding the plasmoid formation precedes the enhancement of electromagnetic turbulence due to shear flows, associated with anomalous momentum transport.
The turbulence is first enhanced around the plasmoid and later, after the plasmoid ejection from the x-line, expands toward the X-line.
\citet{Fermo2012} found using 2D simulations that the electron Kelvin-Helmholtz instability can generate plasmoids at small-scales in the kinetic regime. 
\citet{Markidis2013} simulated a long 3D current sheet, finding evidence for the complex interaction of the resulting plasmoids formed by the tearing instability and bump-on-tail instability together, which generated electron holes and complex electrostatic fluctuations near the plasmoids.
\citet{Nakamura2021} used 2D PIC simulations of a current sheet initially seeded with an ensemble of magnetic field perturbations in order to generate multiple X-points.
The evolution of the system led to a broadband power-law magnetic energy spectrum with a spectral index of -4 below ion scales.
The merging of islands led to a decrease in the reconnection rate and a reduction of the aspect ratio of the electron diffusion region.
As a result, magnetic islands/plasmoids can grow within the electron diffusion region, allowing an (inverse) energy transfer to larger scales.

While the above simulation works identified instabilities and the development of seemingly turbulent dynamics during the nonlinear evolution of those instabilities, they generally did not perform the detailed statistical analyses of the fluctuations typical of turbulence theory. 
\cite{Leonardis2013} analyzed the simulations of \citet{Daughton2011}, finding evidence of intermittency both in the increments of ${\bf B}$ and in ${\bf j}\cdot{\bf E}$, indicative of the turbulent nature of the dynamics. 
\citet{Pucci2017a} analyzed the outflows of 3D reconnection simulations
with a similar setup and parameters as those of \citet{Vapirev2013} and \citet{Lapenta2015}, demonstrating the development of a turbulent cascade and intense dissipation at the boundary between the reconnection outflow and the ambient plasma with turbulence statistics similar to those reported in magnetotail observations. 
Further work explored the dynamics associated with the collision of two reconnection jets in an O-point/magnetic island geometry, showing the development of broadband non-stationary fluctuations between the ion and electron cyclotron frequencies confined to the interaction region between the jets. 
\citet{Munoz2018b} characterized the magnetic spectra in simulations similar to those of \citet{Che2011}, demonstrating the development of broadband kinetic-scale turbulence with typical frequencies between the lower-hybrid and up to the electron-cyclotron frequencies. 
The magnetic spectra steepened as the instabilities evolved, eventually reaching a power-law of $\sim k^{-2.8}$ (typical of observations of kinetic-scale turbulence) once the reconnection rate reached a normalized value of 0.1. 
Interestingly, the reconnection rate is enhanced beyond the typical value of 0.1 as the system evolved further, in conjunction with the spectral slope steepening.
\citet{Zharkova2021} performed a similar analysis to \citet{Munoz2018b} but for a simulation with a more extended current sheet that led to the formation of more plasmoids, emphasising the role of accelerated particles in generating the unstable beam distributions that excite the turbulence. 
In contrast to the previous works, \citet{Adhikari2020} analyzed not a turbulent but initially laminar 2D PIC reconnection simulation, analyzing the spectral properties of the resulting fluctuations.
In steady state, where the reconnection rate was near 0.1 (in normalized units), the fluctuations featured a double power law spectra following $\sim k^{-5/3}$ for scales larger than $d_i$ (fluid scales) and $\sim k^{-8/3}$ for scales between $d_i$ and $d_e$ (kinetic scales).
Similar to the 3D simulations by \citet{Munoz2018b}, \citet{Adhikari2020} found a correlation between the reconnection rate and the energy spectrum, but mainly for wavenumbers near $d_i^{-1}$.
In addition, \citet{Adhikari2020} determined that, while initial energy spectrum associated with the current sheet is highly anisotropic, the anisotropy diminishes as reconnection develops. 
\citet{Lapenta2020a} explored the spectra of the turbulent fluctuations in diverse regions around the main reconnection site using 3D fully kinetic PIC simulations, finding a relationship between the local values of the plasma-$\beta$ and the coupling between plasma and electromagnetic field fluctuations
In the high-$\beta$ regions corresponding to the reconnection outflows, plasma and electromagnetic fluctuations are coupled and turbulent, while, in the low-$\beta$ region corresponding to the reconnection inflow, diffusion region and around the separatrices, the plasma flows appear laminar while the electromagnetic fluctuations appear turbulent. 
Such results potentially suggest that anomalous resistivity and viscosity primarily play a role in the outflow regions.

\section{The Kelvin-Helmholtz Instability \& the Role of Reconnection in the Transition to Turbulence} \label{sec:KHI}
Alongside reconnection, other large-scale structures/instabilities can drive turbulence in a plasma and reconnection can play role as a secondary instability in the initial development of the system into a turbulent state. 
While this scenario has connections with those discussed in Secs.~\ref{sec:TurbulenceDrivenReconnection} and \ref{sec:ReconnectionDrivenTurbulence}, it is worth considering some of its unique features. 
Several numerical studies have looked at the role of reconnection in the destabilization of large-scale initial configurations and the onset of turbulence, noting the apparent non-local energy exchange as reconnection rapidly excites small sub-proton-scale fluctuations \citep{Gingell2017, Franci2017, Manzini2023}. 
One region where the excitation of such a large-scale instability and the subsequent transition to turbulence has been clearly observed in space plasmas is along the strong velocity shear boundary on the flanks of Earth's magnetopause in which the Kelvin-Helmholtz instability (KHI) is well-known to occur. 
Magnetic and velocity shears, such as those found at the magnetopause, coexist in many boundaries in dynamic plasma environments, such as planetary magnetopauses, the heliopause, solar and stellar flares, and astrophysical jets.
While, for sufficiently large magnetic shears, reconnection is expected to be a dominant process, strong velocity shears approaching the Alfv\'en speed based on the sheared component of ${\bf B}$ are expected to suppress instabilities, such as the tearing instability, that are thought to initiate reconnection \citep{Chen1990, Faganello2010} and can excite the KHI.  
Theory and numerical simulations suggest the vortical flow produced by the nonlinear evolution of the KHI can strongly distort and twist ${\bf B}$, inducing secondary reconnection that may contribute to the evolution of the KHI into a turbulent boundary layer.
Recent large-scale simulations further predict that this so-called vortex-induced reconnection (VIR) process can cause mass and energy transfer as efficiently as that caused by regular reconnection induced under large magnetic shears and in-situ observations by various spacecraft have confirmed the evolution of VIR at the magnetopause.

\subsection{Vortex-Induced Reconnection}
\begin{figure*}[h!]
\centering
\includegraphics[width=0.75\textwidth]{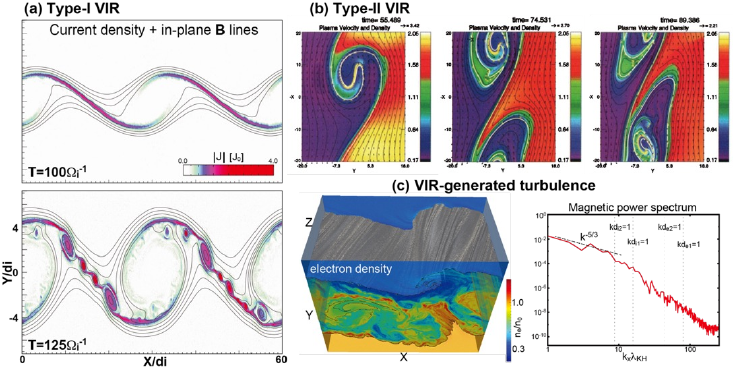}
\caption{(a) Current density and in-plane magnetic field lines of a 2D fully kinetic simulation of Type-I VIR, showing formation of multiple magnetic islands in the compressed current layer \citep[reproduced from][]{Nakamura2013}. 
(b) Plasma density and field lines of a 2-D MHD simulation of Type-II VIR, showing island formation in vortex arms \citep[reproduced from][]{Nykyri2001}.
(c) 3D view of electron density and magnetic power spectrum of a 3D fully kinetic simulation of Type-I VIR, showing generation of turbulence \citep[reproduced from][]{Nakamura2017a}.}
\label{fig:Figure_KH1}
\end{figure*}

Based on linear MHD theories and 2-D two-fluid simulations, \citet{Nakamura2008} summarized two-types of VIR that can be excited in the 2D vortex plane. 
Type-I VIR occurs when pre-existing magnetic shear is locally compressed by the non-linear vortex flow \citep{Pu1990, Knoll2002} as shown in Fig.~\ref{fig:Figure_KH1}a. 
Since Type-I VIR reconnects field lines originally located on different sides of the shear layer, this process can cause rapid plasma mixing across the layer \citep{Nakamura2011, Nakamura2014}. 
In addition, linear theory predicts Type-I VIR is commonly triggered in boundaries where moderate amplitude magnetic and velocity shears co-exist, as often seen at Earth’s magnetopause \citep{Nakamura2006}. 

Type-II VIR is driven in highly rolled-up vortices where the wrapped field lines secondarily form thin current sheets \citep{Nykyri2001, Nakamura2005, Faganello2008} as shown in Fig.~\ref{fig:Figure_KH1}b. 
\citet{Nakamura2005} performed a parameter study suggesting Type-II VIR is triggered when the velocity shear is significantly larger ($\gtrsim5\times$) than the Alfv\'en speed based on the component of ${\bf B}$ along the shear.
Since Type-II VIR divides the vortex and forms magnetic islands penetrating through the vortex layer, this process can also cause efficient plasma transport across the boundary \citep{Nykyri2001}. 
However, since the magnetic topology change in Type-II VIR occurs within a single wrapped field line, the process of Type-II VIR on its own cannot cause plasma mixing in 2D and additional 3D effects, collisionless cross-field diffusion, and/or the coupling with other types of VIR are necessary to enable cross-field mixing and transport. 
Based on numerical simulations, both types of VIR are expected to coexist in the KHI vortex layer \citep{Nakamura2008, Nakamura2013, Karimabadi2013}. 

In these VIR processes, vortical flows strongly compress the current layers down to electron-scales \citep{Nakamura2011}. 
When the length of the compressed current layers, which depends on the size of the parent KH vortex, is sufficiently long compared to electron-scales, multiple plasmoids, with initial sizes comparable to election scales, are observed to form in the compressed current layers for both Type-I and Type-II VIR based on numerical simulations \citep{Nakamura2011, Nakamura2013, Karimabadi2013}, as shown in Fig.~\ref{fig:Figure_KH1}a. 
Reminiscent of the plasmoid instability discussed in Sec.~\ref{sec:ReconnectionDrivenTurbulence}, this tertiary instability contributes to the generation of complex turbulent dynamics within the shear layer through the interaction of magnetic islands from electron to MHD scales. 
In addition, recent 3D kinetic simulations demonstrated that, Type-I VIR can be triggered and evolve over a broad range of oblique angles, which significantly enhances the rate of plasma mixing, as well as the amplitude of the turbulence \citep{Nakamura2013, Nakamura2017a} as shown in Fig.~\ref{fig:Figure_KH1}c. 

While the above studies considered relatively small magnetic shears appropriate to northward interplanetary magnetic field (IMF) conditions at the magnetopause, 3D MHD and kinetic simulations have suggested that, when the magnetic shear is large enough, the turbulent evolution of Type-I VIR quickly disturbs and destroys the vortex structure \citep{Ma2014, Nakamura2020a}. 
On the other hand, more recent 3D kinetic simulations modeling realistic magnetopause conditions under southward IMF showed that when the density jump across the magnetopause is sufficiently large, which would happen more often under southward IMF, the rapid evolution of the LHDI at the compressed current layers quickly diffuses the layers and suppresses Type-I VIR \citep{Nakamura2022a}, although the substructure produced by the vortex-induced LHDI can itself induce small-scale reconnection \citep{Nakamura2022b}. 

\begin{figure*}[h!]
\centering
\includegraphics[width=0.6\textwidth]{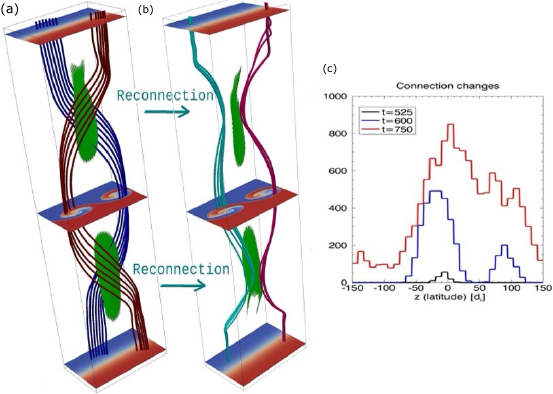}
\caption{(a) 3D view of field line deformation due to differential advection in the symmetric case. 
Blue (magnetospheric) and red (magnetosheath) field lines are initially parallel and twisted by differential advection, creating magnetic shears (green regions). 
(b) Double-reconnected field lines created by ML VIR, with pale blue lines connecting to magnetospheric (blue) plasma at high-latitudes and magnetosheath (red) plasma at the equator \citep[adapted from][]{Faganello2012a}.
(c) Latitude distribution of connection changes (created by reconnection), for different times, in the asymmetric case. 
At t=600, the main peak is due to Type-I VIR, while the secondary peak is due to ML VIR \citep[reproduced from][]{Sisti2019}.}
\label{fig:Figure_KH2}
\end{figure*}

Although most of the above assumed the initial equilibrium varied only in the boundary normal direction, the conditions can also vary in different directions in many realistic situations.
At Earth’s magnetopause, the KHI is unstable at lower latitudes on the magnetopause and stable above and below at higher latitudes \citep{Takagi2006}.
Numerical simulations modeling these conditions have demonstrated the KHI vortex motion twists ${\bf B}$ generating additional magnetic shears in the transition region between the low-latitude unstable and high-latitude stable layers, inducing reconnection at mid-latitudes as illustrated in Fig.~\ref{fig:Figure_KH2}a,b. 
Contrary to Type I VIR, in this mid-latitude reconnection (ML VIR) magnetic shear is created even if the pre-existing magnetic fields are aligned across the boundary. 
Furthermore, since the evolution of the KH vortices and thus ML VIR is symmetric with respect to the equatorial plane, reconnection occurs simultaneously in both hemispheres, creating ``double-reconnected'' field lines topologically connected to the Earth but embedded in the magnetosheath at low-latitudes \citep{Faganello2012a, Faganello2012b, Borgogno2015}. 
The creation of such field lines enhances particle transport across the boundary \citep{Faganello2012b, Borgogno2015}, and can explain the specific entropy increase on the magnetospheric side of the boundary \citep{Johnson2009}.

In the presence of pre-existing magnetic shear, both Type-I VIR and ML VIR occur, with Type-I VIR dominating at low latitudes.
The pre-existing magnetic shear breaks the symmetry in the ML VIR process - gradually enhancing the magnetic shear by differential advection in one hemisphere, while reducing it in the other. 
The evolution of magnetic topology is quite complex, with a dominant Type-I VIR acting close to the equatorial plane, and ML VIR going on in only one hemisphere, as shown in Fig.~\ref{fig:Figure_KH2}c, leading to a less efficient, but still important, production of double-reconnected lines \citep{Sisti2019, Faganello2022}.

\subsection{Observations of Vortex-Induced Reconnection} \label{sec:KH_obs}
\citet{Hasegawa2004} presented evidence of the coexistence of solar wind and magnetospheric ions on the same field lines, suggesting particle transport across the magnetopause, in association with KH waves using {\it Cluster} observations. 
However, magnetic reconnection was initially ruled out as the cause of the plasma transport given the absence of ion reconnection exhaust observations and reconnection-associated D-shaped ion velocity distributions \citep[e.g.,][]{Cowley1982, Fuselier2014} throughout the KH-active interval.
A further analysis of the same event by \citet{Hasegawa2009} provided the first direct  indication that magnetic reconnection was involved with the plasma trasport associated with the KHI on the flanks of Earth's magnetopause. 
High-cadence ${\bf B}$ measurements across the so-called spine-regions between neighboring KH vortices captured several localized current sheets with thicknesses of only a few $d_i$.
One such current sheet showed evidence of bifrucation \citep{Gosling2008} and Alfv\'enic outflow signatures, inferred from the ${\bf E}\times{\bf B}$ drift, in agreement with predictions of Type-I VIR \citep[e.g.,][]{Pu1990, Nakamura2006, Nakamura2008}, although direct measurement of the outflow jet with the particle measurements was not possible. 

${\it MMS}$ provided the high cadence ion measurement capabilities necessary to resolve the ion jets embedded with in the KH vortices.  
\citet{Eriksson2016a} surveyed an extended KH wave-train observed for over an hour at the magnetopause on 8 September 2015, providing the first direct confirmation of ion reconnection exhausts at the quasi-periodic compressed current sheets in the spine region of the KHI associated with Type-I VIR.
In total, reconnection exhausts were found at 22 of the 42 currents sheets with equal probability of encountering reconnection outflows in either direction relative to the x-lines 
Thicknesses normal to the reconnecting current sheets were $4.4\pm1.9 d_i$. 
In addition to outflow signatures, Hall magnetic field perturbations \citep{Eriksson2016a} and particle fluxes across the locally open magnetopause \citep{Li2016, Vernisse2016} were also identified in these events. 
Furthermore, in one case, evidence of the EDR was also identified with two of the {\it MMS} spacecraft ({\it MMS}1 and {\it MMS}2) encountering the ion exhaust, while the other two spacecraft did not but instead encountered signatures indicative of the EDR \citep{Eriksson2016b}. 
In particular, strong parallel electric fields ($E_{||} \sim -15$ mV/m) and enhanced ${\bf j}\cdot{\bf E}^{\prime}\sim8-9$ nW/m$^3$ were identified. 
Since the KHI events under generally northward IMF conditions are expected to produce reconnection at current sheets with relatively low magnetic shear, this event provided one of the first direct measurements of the EDR under strong guide field conditions, with an observed guide field $\sim4\times$ the reconnecting field.

Evidence of ML VIR has also been identified in observations. 
\citet{Faganello2014} reported a short-duration interval of 100–500 eV counter-streaming electrons in {\it THEMIS} observations and interpreted them as magnetosheath electrons accelerated along recently closed field lines at two ML VIR regions, and \citet{Vernisse2016} confirmed a similar signature in {\it MMS} electron measurements in the 8 September 2015 KHI event. 
\citet{Eriksson2021} also reported several short-duration``burst'' of counter-streaming field-aligned ion beams in the 8 September 2015 KHI event. 
The ion velocity distributions were typically ``D-shaped" in phase-space, as commonly associated with magnetopause reconnection \citep{Cowley1982}. 
It was concluded that the counter-streaming ion beams, which were encountered in the leading edge of the vortices in contrast with the Type-I VIR events that were encountered at the trailing edge current sheets, were associated with two nearby ML VIR events above and below the spacecraft, in agreement with predictions of double-reconnected field lines in numerical simulations \citep{Sisti2019}.

\begin{figure*}[h!]
\centering
\includegraphics[width=0.7\textwidth]{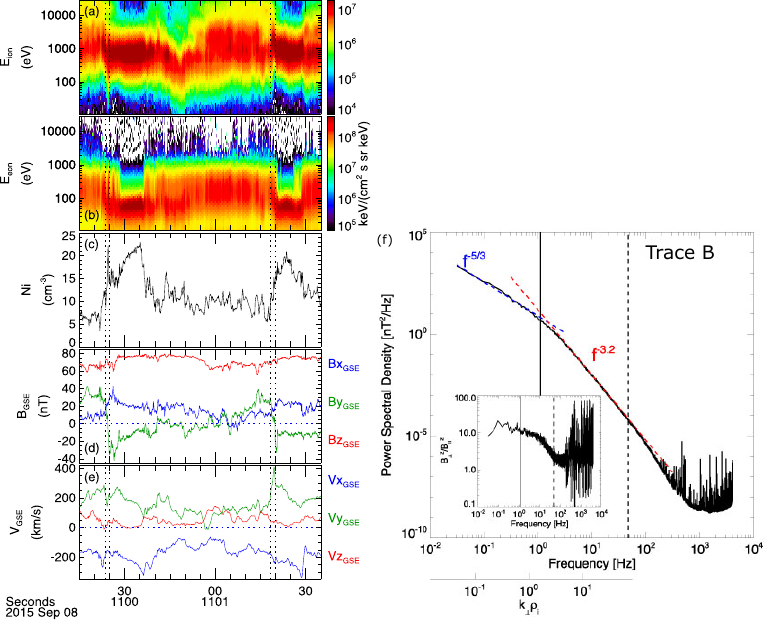}
\caption{Example {\it MMS} observations of two consecutive KH vortices associated with Type-I VIR ion exhausts as highlighted between pairs of vertical dotted lines, showing the (a) omnidirectional ion differential energy flux, (b) omnidirectional electron differential energy flux, (c) ion number density, (d) ${\bf B}$, and (e) ion bulk flow velocity. 
Vector quantities are shown in GSE coordinates. 
(f) Magnetic power spectrum within the KH vortices observed by {\it MMS} on 8 September 2015, which is characterized by a Kolmogorov-like ($k^{-5/3}$) power law at MHD scales \citep[reproduced from][]{Stawarz2016}.}
\label{fig:Figure_KH3}
\end{figure*}

Many other aspects of plasma dynamics have been explored using {\it MMS} observations from the 8 September 2015 KHI event, including several detailed reports on the generation of plasma turbulence and plasma wave activity within the vortices \citep[e.g.,][]{Stawarz2016, Wilder2016, Nakamura2017a, Sturner2018, Sorriso-Valvo2019, Hasegawa2020, Quijia2021}. 
\citet{Stawarz2016} showed that electromagnetic fluctuations observed in the KH vortices were characterized by a Kolmogorov-like power spectrum at large MHD scales (Fig.~\ref{fig:Figure_KH3}b) with signatures of intermittency potentially associated with secondary current sheets formed through the nonlinear KHI development. 
\citet{Hasegawa2020} further investigated the turbulence suggesting the ${\bf B}$ fluctuations in the KH vortices were convective structures associated with interlinked flux tubes generated through 3D turbulent VIR \citep{Nakamura2013, Nakamura2017b}, rather than propagating waves. 
Further analyses have attempted to probe the evolution of KHI turbulence by examining KHI events encountered at different distances along the magnetopause by {\it THEMIS} and {\it Geotail}, although challenges arise from convolving event-to-event variation with the temporal evolution \citep{DiMare2019}.

While the above studies primarily focused on turbulence generation within KH vortices through the lens of KHI and VIR processes, seed perturbations induced by pre-existing magnetosheath turbulence can also act to enhance the KHI growth, and were suggested as an explanation for the apparently rapid onset of turbulent dynamics in the 8 September 2015 KHI event compared to simulations \citep{Nykyri2017, Nakamura2020b}.
Furthermore, for southward IMF, the KHI growth can not only induce reconnection but also form thin density gradient layers at the edge of the KH waves or vortices, which may become unstable to the LHDI \citep{Blasl2022}. 
The resulting LHDI waves or turbulence can in turn cause diffusive plasma mixing across the magnetopause \citep{Nakamura2022b}. 
Thus, there may be an intriguing interplay among turbulence, magnetic reconnection, and MHD and kinetic instabilities, which needs to be further explored.

\section{Stochastic Reconnection} \label{sec:StocasticReconnection} 
A final facet of turbulent reconnection is so-called stochastic reconnection, whereby the wandering of magnetic field lines associated with the turbulent fluctuations plays a dominant role in setting the reconnection rate. 
The ways in which turbulence may impact magnetic reconnection in this manner were discussed by \citet{Matthaeus1986}; however, more complete theoretical treatments were developed by \citet{Lazarian1999} and \citet{Eyink2011}. 
Several reviews discussing the details of this topic have been written over the years \citep[e.g.,][]{Lazarian2012, Lazarian2015, Lazarian2020} and we will refrain from reiterating a detailed discussion of the process here. 
However, in this review we provide a brief overview of the conceptual picture and how it interfaces with the other facets of turbulent reconnection discussed in the previous sections. 

Stochastic reconnection envisions a ``large-scale'' magnetic shear, potentially associated with a pre-existing current sheet embedded within a turbulent environment or a variation in the field orientation associated with larger-scale turbulent fluctuations, that is superposed with small-scale turbulent fluctuations introducing a stochastic wandering of the field lines. 
This field line wandering is akin to Richardson diffusion in hydrodynamic turbulence and, importantly, introduces a rate of diffusion (or perhaps more intuitively dispersion) of the field lines set by the properties of the turbulence, which is independent of the particular non-ideal mechanisms (be that collisional resistivity or collisionless effects) operating in the plasma. 
In this sense, the stochastic reconnection picture may have applications to both the turbulence-driven reconnection and reconnection-driven turbulence scenarios discussed in the previous sections. 

As discussed in \citet{Liu2024}, $\mathcal{R}$ is dictated by the aspect ratio (thickness over length) of the diffusion region. 
Stochastic reconnection allows the thickness of the diffusion region (often referred to as a reconnection layer in this context) to be set by the stochastic field line wandering, resulting in a much wider reconnection layer than otherwise expected. 
In this way, stochastic reconnection is capable of producing a reconnection rate that is both fast and independent of microphysical nonideal effects.
In terms of generalized Ohm's law \citep[see][]{Liu2024}, one can view stochastic reconnection as exploring the effect of the nonlinear contribution of the ideal MHD term ($-\delta{\bf u}\times\delta{\bf B}$) associated with the turbulent fluctuations on the reconnection process \citep[see the discussion of generalized Ohm's law in][]{Eyink2015, Stawarz2021, Lewis2023}.
Stochastic reconnection is subtly different than facilitating reconnection by invoking anomalous resistivity/viscosity (as discussed in Sec.~\ref{sec:ReconnectionDrivenTurbulence_sims}) with regard to the order of averaging that one considers. 
In contrast to anomalous resistivity, where one considers reconnection of an averaged ${\bf B}$, stochastic reconnection considers the cumulative stochastic effect of reconnection of the full field lines.

\citet{Lazarian1999} presented a physical picture in which the resulting wider reconnection outflow was supported by a multitude of small-scale local reconnection events enabled by the ``rough'' perturbed magnetic field line topology of the turbulent fluctuations in the reconnection layer that cumulatively fill the outflow region. 
For a Goldreich-Sridhar-type spectrum, this picture results in an outflow velocity given by
\begin{equation}
V_{outflow} \approx V_{A,inflow}\mathrm{min}\left[\left(\frac{L_x}{\lambda_{inject}}\right)^{1/2},\left(\frac{\lambda_{inject}}{L_x}\right)^{1/2}\right]\mathcal{M}_{A,\lambda_i}^2,
\end{equation}
where $L_x$ is the length of the reconnection layer along the outflow direction, $\lambda_i$ is the injection scale of the turbulent fluctuations (which might be considered as $\lambda_{C,{\bf B}}$), and $\mathcal{M}_{A,\lambda_i}=\delta u_{\lambda_i}/\delta V_{A,\lambda_i}$ is the turbulent Alfv\'en Mach number for fluctuations at the injection scale. 
\citet{Eyink2011} and \citet{Eyink2015} supported this picture in an alternative more mathematically rigorous way by showing that for turbulent velocity fields ${\bf B}$ is only frozen-in to the flow in a stochastic sense even in the limit of vanishing nonideal effects, obtaining equivalent results to those of \citet{Lazarian1999}. 

The clearest evidence for stochastic reconnection comes from numerical simulations. 
\citet{Kowal2009} examined MHD simulations of a reconnecting current layer with different levels of turbulence manually injected into the system, finding $\mathcal{R}$ scaled in a manner similar to the predictions of \citet{Lazarian1999}. 
Further, while in the absence of turbulence $\mathcal{R}$ scaled with resistivity in a manner consistent with Sweet-Parker reconnection, in the presence of turbulence $\mathcal{R}$ was independent of resistivity. 
The structure of the reconnection layer was broadened in the presence of turbulence and made up of many thinner intense current regions, the cumulative effect of which resulted in the overall $\mathcal{R}$ \citep{Kowal2009, Kowal2012, Vishniac2012}. 
Using large MHD turbulence simulations, \citet{Eyink2013} further demonstrated the Richardson-like dispersion behavior of field lines, enabling stochastic breaking of the frozen-in condition.

Clear observational evidence demonstrating stochastic reconnection has not been extensively demonstrated and, in fact, is likely difficult to obtain from the local {\it in situ} measurements. 
\citet{Lalescu2015} reported similarities between solar wind reconnection events and turbulent MHD simulations in which the stochastic field line wandering effect was demonstrated to occur - although it remains to be seen whether such qualitative similarities are unique to the proposed scenario. 
Tangential evidence has also been suggested for the stochastic violation of the frozen-in theorem based on statistical analyses of the Parker spiral \citep{Eyink2015}. 

A key aspect of stochastic reconnection is that it requires the existence of an extended MHD-scale inertial range into which the reconnection dynamics of interest are well coupled - that is, the lengths, widths, and thicknesses of the reconnection layers are within the MHD-scale inertial range. 
In the solar wind, reconnection outflows can be hundreds or even thousands of $d_i$ \citep{Mistry2017} and the length of the x-line (in the direction orthogonal to the quasi-2D reconnection plane) can be $10^4 d_i$ \citep{Phan2006, Phan2009} - both of which are well within the MHD-scale inertial range of solar wind turbulence. 
The quasi-laminar collisionless reconnection viewpoint would typically treat these large length scales, particularly the width of the observed outflows, as being indicative of traversing the outflow far from the x-line \citep[e.g.,][]{Mistry2015b}. 
Stochastic reconnection, on the other hand, would suggest that these large length scales are driven by the stochastic wandering effect. 
In the solar wind, there is some evidence for complex distorted structure of reconnection outflow boundaries from multipoint observations \citep{Mistry2015a}, which may be suggestive of the stochastic reconnection picture (although not necessarily conclusive). 
However, it is generally challenging to tease out direct evidence distinguishing these view points from {\it in situ} measurements. 

This requirement of an extended MHD-scale inertial range may make stochastic reconnection, most relevant to the solar wind, solar corona, and some astrophysical environments with larger dynamical ranges of fluctuations in contrast to the magnetospheric environments such as the magnetosheath, magnetotail, and magnetopause shear layer, which, while present, have more modest MHD-scale inertial ranges. 
Furthermore, it also does not necessarily preclude the existence of localised kinetic-scale reconnecting structures well described by collisionless reconnection dynamics embedded within the turbulent environment. 
As discussed in Secs.~\ref{sec:TurbulenceDrivenReconnection} -- \ref{sec:KHI}, clear evidence of such kinetic reconnection dynamics is found within turbulent environments, including electron-only reconnection embedded within the magnetosheath and electron diffusion regions that appear consistent with quasi-2D collisionless reconnection in turbulent magnetotail events. 
In this light, the further exploration of stochastic reconnection principles in the context of nonlinear collisionless effects, for example the Hall effect or relatedly advection in electron MHD, may be an interesting avenue of research for environments with less extended MHD-scale inertial ranges.

\section{Conclusions} \label{sec:Conclusion} 

As illustrated in this review, the interaction between magnetic reconnection and turbulence is a rich field of study that can be approached from multiple viewpoints, each of which has unique nuances and applications to a variety of plasma environments. 
In a broad sense, the study of turbulence aims to understand the general nonlinear dynamics arising in a fluid or plasma due to the coupled interaction of a vast array of fluctuations that might consist, for example, of a mixture of waves, current sheets, plasmoids/flux ropes, vortices, and other localized structures. 
As a fundamental plasma process that can occur in a diverse range of settings, magnetic reconnection events have the potential to be one of these nonlinear structures, which can act to mediate the nonlinear interaction of magnetic structures and facilitate energy dissipation both in systems that are in a fully-developed turbulent state or that are developing into one. 
In other configurations, where magnetic reconnection is the primary process acting at a system scale current sheet, secondary processes associated with magnetic reconnection can act to excite turbulent fluctuations in the plasma, which mediate the partition of energy released by the reconnection event. 
Furthermore, the presence of complex turbulent fluctuations can also potentially impact how magnetic reconnection proceeds, for example, by enhancing the reconnection rate through the action of anomalous resistivity/viscosity near the x-line or through stochastic field line wandering. 

While these various aspect of turbulent reconnection have been examined in a wide array of theoretical, numerical, and observational studies over the past several decades, recent high-resolution spacecraft measurements, in particular from the {\it Magnetospheric Multiscale} mission, have ushered in a new era of investigating this complex topic. 
Analyses enabled by these measurements have provided new insights ad observational constraints that have spurred on a range of new theoretical and numerical investigations.
Several areas of recent progress that are of particular note and that have been highlighted in this review are:
\begin{enumerate}
\itemsep=1ex

\item Recent observations of turbulence-driven magnetic reconnection in the transition region and downstream magnetosheath of Earth's bow shock. 
Systematic surveys of small-scale reconnection events in this environment have allowed statistical examinations of how the turbulent dynamics influence the nature of the magnetic reconnection events and the potential role that magnetic reconnection plays in the turbulent dynamics. 
Observations in this environment have revealed that so-called electron-only magnetic reconnection can occur when the correlation length of the turbulent magnetic field fluctuations sufficiently limits the extent of the reconnecting current sheets. 
This was a novel discovery that had not been extensively explored prior to the {\it MMS} observations, which highlights how the properties of the large-scale turbulent fluctuations can shape the dissipative processes operating in the plasma and may have implications for how the energy dissipated by turbulence is partitioned. 

\item Detailed observations of turbulent x-lines and bursty bulk flow braking regions in Earth's magnetotail. 
These regions have revealed how turbulence facilitates the energization of high-energy non-thermal particles, how turbulence mediates the re-partitioning of energy released by magnetic reconnection both through spontaneously generated turbulence and the interaction with the surrounding environment, and demonstrated that identifiable electron diffusion region signatures, reminiscent of the quasi-laminar picture, can be identified even amidst the complex turbulent fluctuations.

\item New observations of the Kelvin-Helmholtz instability on Earth's magnetopause that have both revealed clear evidence of the various forms of vortex-induced magnetic reconnection and the coupled evolution of the instability into a turbulent boundary layer, which can play a key role in the transport of mass, momentum, and energy across the magnetopause and, potentially, other velocity shear boundaries throughout the Universe. 

\end{enumerate} 

Despite this recent progress a variety of open questions remain. 
The systematic identification of turbulence-driven reconnection events allows the estimation of the extent to which magnetic reconnection contributes to turbulent dissipation. 
However, more refined analyses require further constraints on the reconnection rate and particle heating/energisation associated with magnetic reconnection -- in particular, in terms of the contrast between electron-only and ion-coupled reconnection.
Both these factors could, in principle, be further examined using numerical simulations either of fully turbulent domains or well designed numerical experiments using idealised configurations. 

Another key area of advancement will be in adapting our understanding of turbulence-driven reconnection to other astrophysical environments.
Addressing this point requires further understanding of the configuration of reconnecting current sheets (e.g., whether they are fragmented, ``rolled-up'', or otherwise deformed) and how this may depend on the Reynolds number/scale separation in the system, as well as how the prevalence of turbulence-driven magnetic reconnection is impacted by the driving, fluctuation characteristics, and ambient plasma conditions. 
Such factors may have an influence on the number of reconnection events per unit volume in the turbulent domain and the extent to which electron-only reconnection and ion-coupled reconnection are important in different systems.  
A useful path to pursue in this regard may be in contrasting turbulence-driven reconnection across the different turbulent environments that we have access to in near-Earth space, such as Earth's magnetosheath and the solar wind -- particularly given the extreme contrast in effective scale separation between the driving and kinetic scales between these two environments. 
However, a key component of such studies would need to involve clearly assessing the extent to which reconnection events in the solar wind are related to locally generated turbulent structure rather than large-scale solar wind structure. 
Upcoming missions, such as NASA's {\it HelioSwarm} \citep{Klein2023} which aims to specifically target the multi-scale nature of solar wind turbulence with a large swarm of spacecraft, may aid in assessing the role of turbulence-driven reconnection in the solar wind, but it may also require new technology and mission concepts capable of probing the electron scales in the solar wind to determine the extent to which electron-only reconnection may be relevant at fragmented current sheets in the solar wind \citep{Verscharen2022}. 
Numerical studies will also likely be a key component of probing how such dynamics may change in more exotic environments that are not directly accessible with spacecraft observations.
Given the challenge in identifying 3D magnetic reconnection events in the complex magnetic environments of turbulent plasmas, machine learning or partially machine learning based approaches to systematically identifying turbulence-driven magnetic reconnection events may prove to be a useful tool. 

The recent theoretical and numerical work on reconnection-mediated turbulence, which explore the potential role of reconnection in mediating the nonlinear interactions in turbulent plasmas, have also presented intriguing results that still stand to be observationally confirmed. 
However, further observational tests that go beyond strictly spectral slope based analyses may be necessary to unambiguously confirm the presence of this regime given the similarity of the spectral predictions to commonly observed kinetic-scale turbulent spectra that can be predicted by other means. 

In the context of turbulence-driven reconnection, key areas of future research will likely come from developing a further understanding of how the turbulent processes associated with magnetic reconnection couple into and govern the energy transport of the global system. 
For Earth's magnetotail, such work could involve clear and systematic surveys of the extent to which the turbulent fluctuations accelerate energetic particles particularly in the braking region where they may act to seed the energetic particles in the radiation belts, exploring the wave radiation associated with confined regions of turbulence that may be generated in reconnection outflows, and systematically quantifying how the partition of energy evolves with both time and distance from the x-line in turbulent reconnection outflows. 
As well as having relevance to the magnetotail, such studies may also be relevant to reconnection in solar coronal loops, astrophysical jets, and possibly magnetopause reconnection. 
The impact of anomalous resistivity/viscosity and stochastic field line wandering on the magnetic reconnection rate, which is particularly useful for its application in parameterising the effects of turbulence in astrophysical environments, still require clear and unambiguous observational confirmation. 
The potential impacts of turbulent inflow conditions on system-scale reconnection events, such as the impact of turbulent magnetosheath flow on magnetopause reconnection, may also be an interesting and related avenue of research.

Given the recent advances in our understanding of the interplay between turbulence and magnetic reconnection and further upcoming and potential missions such as NASA's {\it HelioSwarm} \citep{Klein2023} and {\it Plasma Observatory} recently proposed to ESA \citep{Retino2022} that specifically target turbulence and cross-scale coupling in the solar wind and Earth's magnetosphere, there is a bright future for the further development of our understanding of this complex problem which stands to unlock new insights into the role of magnetic reconnection throughout the Universe.

\begin{acknowledgements}
JES is supported by the Royal Society University Research Fellowship URF\textbackslash R1\textbackslash 201286. 
R.B. was supported in part by the MMS Early Career Award NASA Grant No.~80NSSC21K1458 and NASA Grant No.~80NSSC21K0739
The authors thank the International Space Science Institute for their support of the ``Magnetic Reconnection: Explosive Energy Conversion in Space Plasmas" workshop through which the contributions to this collection were coordinated and prepared. 
\end{acknowledgements}

\section*{Conflict of interest}
The authors have no conflicts of interest to declare.

\appendix
\section{Taylor's Hypothesis}\label{app:taylor}
Taylor's ``frozen-in flow" hypothesis has been widely used to study the properties of space plasma fluctuations~\citep{Taylor1938PRSLA}. 
The hypothesis states that the advection of the small-scale fluctuations over a measurement point by the large-scale background flow occurs faster than any significant dynamical evolution of those fluctuations. 
Therefore, a time series measured at a single point in space can be interpreted as a spatial sample through the system. 
Using Taylor's hypothesis, $\boldsymbol{\ell}$ is given by $\boldsymbol{\ell} = {\bf U}_0 \tau$, where $\tau$ is a temporal lag. 
Similarly, the spacecraft-frame frequency $(f)$ spectrum can be interpreted as a $k$ spectrum with the relation $k = 2\pi f/U_0$, although the spectrum computed this way should be interpreted as a reduced spectrum averaged onto the direction of ${\bf U}_0$ \citep{Horbury2012}. 
Taylor's hypothesis is satisfied for MHD scales when $U_0 \gg V_A$ and $U_0 \gg \delta u$. 
This assumption is typically well satisfied in the solar wind, where background flows are super-Alfv\'enic; however, in other environments, such as those found in the magnetosphere, the validity is less clear. 
Furthermore, at sub-proton-scales, the validity may be called into question due to the faster phase speeds at those scales. 

\begin{figure*}[h!]
\centering
\includegraphics[width=0.7\textwidth]{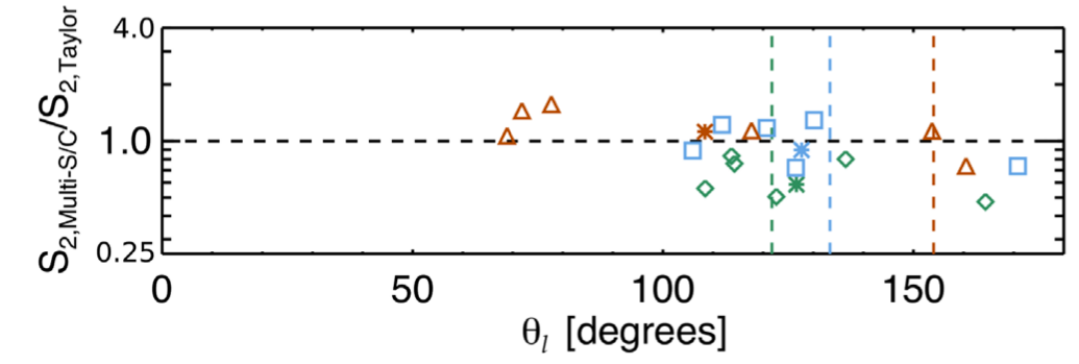}
\caption{Ratio of multi-spacecraft to single-spacecraft estimates of $S_2^{\bf B}\left(\boldsymbol{\ell}\right)$ as a function of $\theta_{\ell}$ for several intervals of magnetosheath turbulence measured by {\it MMS} (blue squares, red triangles, green diamonds). 
The six points for each interval correspond to the six spacecraft pairs in the formation. 
Averages of the six points are marked with asterisks, and vertical dashed lines mark the ${\bf U}_0$ direction~\citep[reproduced from][]{stawarz2019}.}
\label{fig:Figure_Taylor}
\end{figure*}

Using multi-spacecraft missions, such as {\it MMS} or {\it Cluster}, the Taylor hypothesis can be tested at the scale of the multi-spacecraft formation by comparing single-spacecraft statistics (e.g., structure functions or correlation functions) with their multi-point counterparts computed from pairs of spacecraft. 
For anisotropic turbulence, as expected in the presence of a strong ${\bf B}_0$, the correspondence between single- and multi-spacecraft statistics at a given scale may only occur along the direction of ${\bf U}_0$; however, the dependence on the angle between the lag and ${\bf B}_0$ ($\theta_{\ell}$) may provide insight into the validity of the Taylor hypothesis. 
\citet{stawarz2019} compared single- and multi-spacecraft estimates of $S_2^{\bf B}\left(\boldsymbol{\ell}\right)$ for {\it MMS} observations of turbulence in the magnetosheath, as shown in Fig.~\ref{fig:Figure_Taylor}. 
For the super-Alfv\'enic flows (orange triangle and blue sqaures), good agreement is found for both intervals, with no particular dependence on $\theta$, demonstrating the Taylor hypothesis can be valid even down to the small ($\sim 6$km) separations of the {\it MMS} formation, which are much smaller than the proton scales, during these intervals  
In contrast, the green diamonds display a systematic overestimate of the single-spacecraft estimates relative to the multi-spacecraft estimates for all six spacecraft pairs, indicating that the Taylor hypothesis may not work well during this interval. 
\citet{Stawarz2022} performed a similar analysis across a range of turbulent intervals in the magnetosheath, demonstrating a clear signature of anisotropy in the turbulent fluctuations as a function of $\theta_{\ell}$ with increasing strength for $\delta b_{rms}/B_0<1$.
Good agreement was found between single- and multi-point statistics when $\theta_{\ell}$ was equivalent to the angle between ${\bf B}_0$ and ${\bf U}_0$ for intervals where the Taylor hypothesis was valid.
Other similar analyses have been performed by across the magnetosheath and other regions of the magnetosphere \citep[e.g.,][]{Chasapis2017ApJL, Parashar2018PRL, Bandyopadhyay2020MNRAS}.

\bibliographystyle{spbasic} 

\end{document}